\documentclass[reprint,prl,superscriptaddress]{revtex4-1}
\usepackage{graphicx,subfigure,morefloats,multirow}
\usepackage{amsmath,array,amssymb,xparse} 
\usepackage{hyperref}
\usepackage{booktabs}

\usepackage{color}
\usepackage{siunitx}
\usepackage{array}
\usepackage{upgreek}
\usepackage{empheq}

\begin{document}

\title{Predicting Complex Non-spherical Instability Shapes of Inertial Cavitation Bubbles in Viscoelastic Soft Matter}  

\author{Jin Yang}
\affiliation{Department of Mechanical Engineering, University of Wisconsin-Madison}

\author{Anastasia Tzoumaka}%
\affiliation{School of Engineering, Brown University}

\author{Kazuya Murakami}%
\affiliation{Department of Mechanical Engineering, University of Michigan, Ann Arbor}

\author{Eric Johnsen}%
\affiliation{Department of Mechanical Engineering, University of Michigan, Ann Arbor}

\author{David L. Henann}%
\affiliation{School of Engineering, Brown University}

\author{Christian Franck} \email{cfranck@wisc.edu; Corresponding author}
\affiliation{Department of Mechanical Engineering, University of Wisconsin-Madison}
 
\date{\today} 
%%%%%%%%%%%%%%%%%%%%%%%%%%%%%%%%%%%%%%%%%%%%%%%%%%%
\begin{abstract}
Inertial cavitation in soft matter is an important phenomenon featured in a wide array of biological and engineering processes. Recent advances in experimental, theoretical, and numerical techniques have provided access into a world full of nonlinear physics, yet most of our quantitative understanding to date has been centered on a spherically symmetric description of the cavitation process. %Flip the model ahead of the experiments to emphasize the model more
However, cavitation bubble growth and collapse rarely occur in a perfectly symmetrical fashion, particularly in soft materials. % due to material microscopic heterogeneity, non-symmetric boundary and macroscopic forces. 
Predicting the onset of dynamically arising, non-spherical instabilities has remained a significant, unresolved challenge in part due to the additional constitutive complexities introduced by the surrounding nonlinear viscoelastic solid. Here, we provide a new theoretical model capable of accurately predicting the onset  of  non-spherical instability shapes of a bubble in a soft material by explicitly accounting for all pertinent nonlinear interactions between the fluid-like cavitation bubble and the solid-like surroundings. Comparison against  high-resolution experimental images from laser-induced cavitation events in a polyacrylamide (PA) hydrogel show excellent agreement. Interestingly, and consistent with experimental findings, our model predicts the emergence of various dynamic instability shapes for hoop stretch ratios greater than one in contrast to most quasi-static investigations. 
Our new theoretical framework 
%Being able to predict the onset and evolution of various instability shapes and their associated material strain fields from common experimental data 
not only provides unprecedented insight into the cavitation dynamics in a soft solid, but it also provides a  quantitative means of interpreting bubble dynamics relevant to a wide array of engineering and medical applications as well as natural phenomena.    
\end{abstract}


\maketitle
%\tableofcontents

%%%%%%%%%%%%%%%%%%%%%%%%%%%%%%%%%%%%%%%%%%%%%%%%%%%%%%
%\section{Introduction}\label{sec:intro}
%%%%%%%%%%%%%%%%%%%%%%%%%%%%%%%%%%%%%%%%%%%%%%%%%%%%%%

Cavitation is the process whereby vapor cavities, or bubbles, are produced in a liquid or solid due to pressure differences. For inertially generated bubbles, the resulting dynamics are highly transient and often nonlinear.  While models have been developed to successfully describe cavitation dynamics in water, the extension of the theoretical foundation of these models to soft materials is challenging due to their unique deformation and failure mechanisms, especially under extreme loading conditions (large strain magnitudes and high strain rates). In particular, the nature of the physical and chemical composition of many soft polymers gives rise to highly nonlinear elastic and viscoelastic macroscale material behavior. 
%From an application perspective, being able to quantitatively describe the soft matter interaction of inertial cavitation bubbles is a crucial step in the controlled and safe administration of laser eye surgeries \cite{cherian2008pulsed}, histotripsy \cite{Blomley1222,Vlaisavljevich2016,mancia2019modeling}, and targeted treatment of cancer cells \cite{brennen2015cavitation,Mittelstein2020apl,Schibber2020prsa}, just to name a few \jin{examples}.  

Current prediction of inertial cavitation dynamics usually relies on the approximation that the cavitation process remains nominally spherically symmetric through its life cycle, allowing for the use of the classical Rayleigh-Plesset \cite{plesset1954jap,prosperetti1977qam,luo2020scirep,barney2020pnas,saintmichel2020cocis} or Keller-Miksis \cite{keller1980bubble,estrada2018jmps,yang2020eml} modeling approaches. However, recent experiments and numerical simulations show that cavitation bubble growth and collapse rarely occur in a perfectly spherically symmetric fashion even in nominally homogeneous and isotropic soft matter \cite{brenner1995prl,hamaguchi2015pof,shaw2017pof,guedra2018jfm,saintmichel2020softmatter}. 
Predicting the onset of non-spherical deformation arising due to instabilities during inertial cavitation in soft viscoelastic materials is an important and fundamental problem with significant implications across many applications. For example, non-spherical instabilities may lead to strain-localization and subsequent material damage near the bubble wall as well as other important physical, chemical, and biological outcomes, including sonochemistry \cite{yoshikawa2014chemsocrev,mettin2015sonochem}, sonoluminesence \cite{brenner1995prl}, local plasticity and fracture, %of surrounding material near the bubble wall \cite{movahed2016jasa}, and the formation of high-speed liquid jets capable of causing significant structural damage \cite{johnsen_colonius_2009,brujan_nahen_schmidt_vogel_2001,supponen_obreschkow_tinguely_kobel_dorsaz_farhat_2016} 
and/or tissue/cell dysfunctions in biological materials \cite{brennen2015cavitation}.

% ----- A quick summary of our novelty -----
Documentation of cavitation-related instabilities in liquids has a well-established history in the fluid mechanics community, including detailed descriptions of classical instability phenomena, such as the \textit{Rayleigh-Taylor} (RT) instability \cite{taylor1950RTinstab}, or \textit{parametric} instabilities,  which arise due to the accumulation of non-spherical perturbations over many oscillation periods \cite{hamaguchi2015pof,saintmichel2020softmatter,murakami2020us,gaudron2020jmps}. However, predicting the onset of non-spherical instabilities during inertial cavitation in soft materials is still in its infancy, largely due to the additional complexities arising from the intrinsic coupling of the bubble dynamics to a nonlinear, viscoelastic solid. Cavitation in a solid material can present different and perhaps more complex (non-spherical) instability patterns compared to a fluid, including wrinkles, creases, and folds. In addition, parametric instabilities, tend to occur earlier during the cavitation expansion-collapse cycles in a solid material when compared to a fluid. Mathematically, the constitutive laws of hyperelastic solids are  typically expressed using a Lagrangian description based on the reference configuration, while the cavitation dynamics are typically described in the current, deformed configuration, which must be accounted for when theoretically describing and predicting the onset and evolution of inertial cavitation instability patterns within a soft solid.

%For example, (i) although the surrounding fluid and solid materials can both be assumed to be nearly incompressible during an inertial cavitation event \cite{estrada2018jmps,yang2020eml}, they follow different mathematical rules: the surrounding flow velocity is divergence free in a fluid material \cite{murakami2020us}, whereas the determinant of the deformation gradient tensor equals to one in a solid material. Thus, non-spherical perturbations in the kinematic field of a solid or fluid will have a different mathematical starting point, or ansatz; (ii) cavitation in a solid material can present distinctively different, and perhaps more complex (non-spherical) instability patterns as compared to a fluid including wrinkles, creases, and folds. In addition, parametric instabilities, tend to occur earlier during the cavitation expansion-collapse cycles in a solid material when compared to its liquid counterpart; (iii) the constitutive laws of hyperelastic solids are usually expressed by a Lagrangian description based on the reference configuration, while the cavitation dynamics are typically modeled based on the current, deformed configuration. These are just some examples of the intricacies that one needs to consider when attempting to theoretically describe and predict the onset and evolution of inertial cavitation instability patterns within a soft solid.

To address these challenges, we present a new theoretical framework that, based on a first-order incremental perturbation analysis, is able to  accurately capture and predict the evolution of complex deformation modes observed in dynamic cavitation events. By comparing our theoretical predictions against recent experimental observations of various non-spherical bubble shapes, we find that our model is in excellent quantitative and qualitative agreement with the experimental measurements. 

%%%%%%%%%%%%%%%%%%%%%%%%%%%%%%%%%%%%%%%%%% END OF INTRO  %%%%%%%%%%%%%%%%%%%%%%%%%%%%%%%%%%%%%%%%%%%%%%%

%%%%%%%%%%%%%%%%%%%%%%%%%%%%%%%%%%%%%%%%%%%%%%
%\section{Experiments}\label{sec:exp}
%%%%%%%%%%%%%%%%%%%%%%%%%%%%%%%%%%%%%%%%%%%%%%
%\subsection{Experimental Setup}\label{sec:cav_setup}

%%%%%%%%%%%%%%%%%%%%%%%%%%%%%%%%%%%%%%%%%%%%%%
\begin{figure}[!t]
\begin{center} 
\includegraphics[width=\columnwidth]{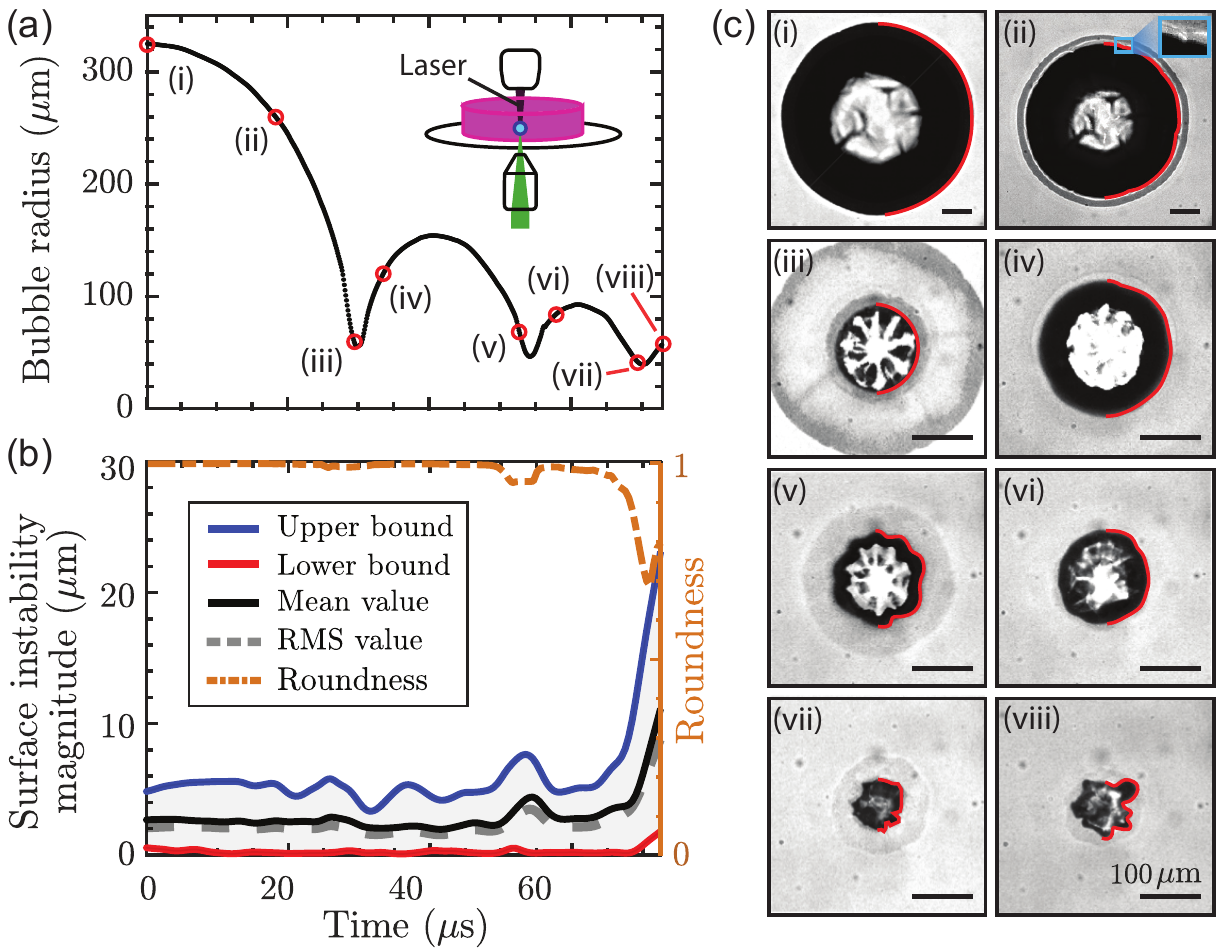} 
\end{center}
\caption{Spatiotemporally recorded bubble dynamics via high-speed videography \cite{yang2020eml}. (a)\,A representative bubble radius vs.\ time curve for a laser-induced cavitation bubble inside a PA gel. (b)\,Quantitative measurements of the magnitude of bubble surface instabilities. (c)\,Various types of non-spherical bubble shapes corresponding to different time-points (red circles) in (a). Red dashed lines are overlaid onto original images to visualize the shape of the bubble wall. Inset in (c-ii) is the zoomed-in non-spherical shape instability. Data from \cite{yang2020eml}.} \label{fig:exp4}
\end{figure}

In order to compare and assess the accuracy of our theoretical predictions against appropriate experiments, we take advantage of several existing high-resolution data sets of laser-induced cavitation in polyacrylamide (PA) hydrogels \cite{yang2020eml}. 
%Here, several single bubble inertial cavitation events \jin{were} generated with a single 6 ns pulse  Q-switched Nd:YAG Minilite II (Continuum, Milpitas, CA) laser platform producing bubbles with average maximum and equilibrium \jin{bubble} radii of 330.3\,$\pm$\,30.4\,$\upmu$m, and 46.5\,$\pm$\,4.3\,$\upmu$m, respectively. All cavitation dynamics were recorded at 2 million frames per second with a Kirana5M high speed camera (Specialized Imaging, Pitstone, United Kingdom), while full-field illumination was provided by synchronously triggered illumination pulses from a SI-LUX640 laser illumination system (Specialized Imaging, Pitstone, United Kingdom) \cite{yang2020eml}. 
A representative bubble radius vs.\ time curve is shown in Fig.\,\ref{fig:exp4}(a), plotted starting at the time-point of the maximum bubble radius $R_{\text{max}}$. %While a previous investigation of this data set only examined the data until the first collapse point \cite{yang2020eml}, %with the goal of establishing a critical, "violent" Mach number and material property determination scheme 
%our analysis here focuses instead on the spatiotemporal details of the shape evolution of the bubble wall through the entire bubble oscillation lifecycle with the goal to develop a new, predictive theoretical framework. 
Figure\,\ref{fig:exp4}(c) depicts various non-spherical bubble shapes at  eight time-points during the expansion and collapse cycle of the cavitation dynamics, as indicated by red circles in Fig.\,\ref{fig:exp4}(a). The bubble shapes in frames (i,ii,iv,vi) show nearly spherical bubbles, while frames (v,vii,viii) present wrinkling, creasing, and stellate instabilities near the bubble wall, respectively.  From these experimental observations, we quantify the upper and lower bounds, mean, and root-mean-square (RMS) values of the magnitude of these bubble surface instabilities \footnote{Upper and lower bounds, mean, and RMS values of the magnitude of bubble surface instabilities are summarized in Supplementary Material S2.}, as well as the bubble roundness (4$\pi\times$Area/Perimeter$^2$) (see\ Fig.\,\ref{fig:exp4}(b)). 

%%%%%%%%%%%%%%%%%%%%%%%%%%%%%%%%%%%%%%%%%%%%%%
\begin{figure}[!t]
\begin{center} 
\includegraphics[width=.48 \textwidth]{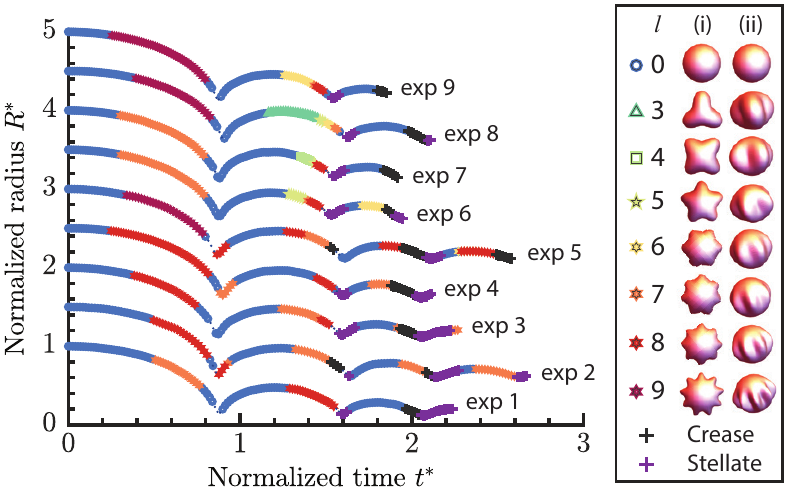} 
\end{center}
\caption{Summary of various instability patterns in nine individual cavitation events. Experimental data sets 2-9 are purposefully offset vertically for easier viewing of each curve. Right inset: (i) Top and (ii) side views of different  instability modes.} \label{fig:expAll}
\end{figure}
%%%%%%%%%%%%%%%%%%%%%%%%%%%%%%%%%%%%%%%%%%%%%%

Next, we non-dimensionalize all bubble radius vs.\ time curves ($t^{*}$\,$=$\,$t {\small \sqrt{p_{\infty}/\rho}}$\,$ / $\,$R_{\text{max}}$, $R^{*}$\,$=$\,$R$\,$/$\,$R_{\text{max}}$, where $p_\infty$ and $\rho$ are defined after Eq.~\eqref{Eq:RayleighPlesset}), and present our findings for the various instability patterns for nine individual experiments in Fig.\,\ref{fig:expAll}. Experimental data sets 2-9 are purposefully offset vertically for easier viewing of each curve. Top and side views of different instability modes (mode shape number $l$ later defined in Eq.\,(\ref{eq:inc_def_disp})) are summarized in the right inset.  The instability pattern within each frame is directly extracted from that frame. Creases are distinguishable from wrinkles by their innate surface folds and significant local curvature in the formed creases (Fig.\,\ref{fig:exp4}(c-vii)) \cite{diab2013prsa}, whereas the emergence of stellate patterns are typically observed right after the occurrence of local creases (Fig.\,\ref{fig:exp4}(c-viii)). By conducting a careful qualitative shape analysis, we find the first appearance of non-spherical bubble shapes near the first, often violent (local Mach number exceeding 0.08), collapse, predominantly featuring shapes similar to the spherical harmonic function of mode shape 8 (Fig.\,\ref{fig:exp4}(c-v)), while subsequent collapse points show the emergence of other mode shapes  (e.g., spherical harmonic function mode shapes 6 and 7) along with mode shape 8.

%%%%%%%%%%%%%%%%%%%%%%%%%%%%%%%%%%%%%%%%%%%%%%
%\section{Theoretical model}\label{sec:theory}
%%%%%%%%%%%%%%%%%%%%%%%%%%%%%%%%%%%%%%%%%%%%%%
% ----- Theoretical tool: inc-def-analysis -----
To quantify the critical condition for the onset of non-spherical instabilities  associated with each mode shape, we develop a new theoretical framework based on a perturbation approach of the spherically symmetric Rayleigh-Plesset governing equation \cite{plesset1949dynamics}, in which the time-dependent bubble radius of the base state $R(t)$ is governed by \cite{plesset1949dynamics,estrada2018jmps,yang2020eml,murakami2020us}:   %the modified Rayleigh-Plesset equation
\begin{equation}\label{Eq:RayleighPlesset}
R R_{,tt} + \frac{3}{2} R_{,t}^2 = \frac{1}{\rho} \left( p_b - p_{\infty} + S - \frac{2 \gamma}{R} \right),
\end{equation}
where  $(\cdot)_{,t}$ denotes the derivative with respect to time; $\rho$ is the mass density of the surrounding viscoelastic material, which is assumed to be nearly constant; $\gamma$ is the surface tension between the gaseous bubble phase and the surrounding medium; $p_b$ is the internal bubble pressure; $p_{\infty}$ is the far-field pressure, which is assumed to be atmospheric as there are no external driving forces; and $S$ is given through the integral of deviatoric Cauchy stress components over the surroundings. Consistent with our previous work, we describe the surrounding soft material as a nonlinear strain-stiffening Kelvin-Voigt (qKV) viscoelastic material \cite{yang2020eml} with a quadratic-law strain energy density function $W$ given by
\begin{equation}
W = \frac{G}{2}  \left[(I_1 - 3) + \frac{\alpha}{2}(I_1-3)^2  \right],   \label{eq:W_quad}
\end{equation} 
where $I_1$ is the first invariant of the right Cauchy-Green deformation tensor, $G$ is the ground-state shear modulus, and $\alpha$ is a dimensionless material parameter characterizing large deformation strain stiffening effects \cite{yang2020eml}. 
%We find that for each normal mode of the spherical harmonic function, the equations of motion reduce to two sets of decoupled evolution equations: the mean bubble radius and the non-spherical bubble amplitude \cite{plesset1954jap,prosperetti1977qam,murakami2020us,gaudron2020jmps}.  
%
The Newtonian viscosity $\mu$ in the qKV model is assumed to be constant and typically has a value of $O(10^{-3})$\,$\sim$\,$O(10^{-1})$ Pa$\cdot$s for water-based hydrogels and biomaterials \cite{yang2020eml,mancia2020amr}. The stress integral $S$ then takes the form 
\begin{equation}
\small
\begin{aligned}
 S =& \frac{(3 \alpha - 1)G}{2}  \left[ 5  - \left( \frac{R_{0}}{R} \right)^4  - \frac{4 R_{0}}{R} \right] - \frac{4 \mu \dot{R}}{R} + \\
&  2 \alpha G  \left[\frac{27}{40} + \frac{1}{8} \left( \frac{{R}_{0}}{R} \right)^8 + \frac{1}{5}\left( \frac{{R}_{0}}{R} \right)^5 + \left( \frac{{R}_{0}}{R} \right)^2 - \frac{2R}{R_{0}}  \right]  
\end{aligned}  \label{eq:S_quad} 
\end{equation}  
where $R_{0}$ is the undeformed bubble radius. %, i.e., $I_1 = \lambda_1^2 + \lambda_2^2 + \lambda_3^2$ and $\lambda_i$ are principal stretches. $G$ is the ground-state shear modulus from quasi-static measurements, and $\alpha$ is an explicit parameter to describe the strain stiffening effects \cite{yang2020eml}.   
We model the inside of the bubble as a two-phase mixture consisting of water vapor and non-condensible gas, which is homobaric and follows the ideal gas law. %Mass and heat transfer of the bubble internal contents gas are assumed to obey Fick's law and Fourier's law, respectively {\color{red} (see Supplementary Material Section S3)}. 
We neglect temperature changes at the bubble wall and assume there is no mass diffusion of non-condensible gas across the bubble wall since both the heat and mass diffusion processes across the bubble wall are much slower than the bubble dynamics. 
Simulations are initiated when the bubble attains its maximum radius $R_{\text{max}}$ thus avoiding the need to account for non-equilibrium nucleation and growth dynamics and beginning the simulations at thermodynamic equilibrium \cite{barajas2017effects,estrada2018jmps}.

%%%%%%%%%%%%%%%%%%%%%%%%%%%%%%%%%%%%%%%%%%%% 
%\subsection{Incremental deformation perturbation analysis}\label{sec:inc_def}
%Here we consider a spherical cavitated bubble with a reference configuration $\mathcal{B}_{0}(\mathbf{X})$ or $\mathcal{B}_{0}(r_0,\theta_0,\phi_0)$:\,$\lbrace R_{0} \leqslant r_0 \leqslant \infty, 0 \leqslant  \theta_0 \leqslant \pi, 0 \leqslant \phi_0 \leqslant 2 \pi \rbrace $, and the current deformed configuration is $\mathcal{B}(\mathbf{x})$ or $\mathcal{B}(r,\theta,\phi)$:\,$\lbrace R \leqslant r \leqslant \infty, 0 \leqslant \theta \leqslant \pi, 0 \leqslant \phi \leqslant 2\pi \rbrace$, see Fig.\,\ref{fig:sph_coord}, where $\lbrace r_0, r \rbrace$ represent radial axes, $\lbrace \theta_0, \theta \rbrace$
%and $\lbrace \phi_0, \phi \rbrace$ are azimuthal and polar angle directions. We further define radial stretch ratio as $ \lambda = r/r_0 $.

Next, based on the spherically symmetric Rayleigh-Plesset governing equation \eqref{Eq:RayleighPlesset},  we consider the following perturbation of the displacement field using spherical harmonic basis functions \footnote{Derivation of the ansatz for the perturbed deformation is summarized in Supplementary Material S3.}: 
\begin{equation}
\left\lbrace 
\begin{aligned}
\tilde{u}_r &= \frac{a(t) R^2}{r^2} Y_{l}^{m} (\theta,\phi) \quad (l \geqslant |m| >0) \\
\tilde{u}_{\theta} &= 0 \\
\tilde{u}_{\phi} &= 0 
\end{aligned}
\right. \label{eq:inc_def_disp}
\end{equation}
where $\lbrace r, \theta, \phi \rbrace$ are radial, polar, and azimuthal angular coordinates; %The overdot denotes the perturbed increment based on the current configuration. 
$Y_{l}^{m} (\theta,\phi)$
%  := $\text{cos}(m \phi) P_l^m (\text{cos}\theta)$ 
is the normalized $(l,m)^{th}$ spherical harmonic function; %, and $l,m$ are the mode order and degree respectively.  
%$P_l^m (\text{cos}\theta)$ is the associated Legendre polynomial function. 
and $a(t)$ is the time-dependent perturbation magnitude at the bubble wall. %; $h(r)$ is the magnitude of the perturbed hydrostatic pressure, $\dot{p}$, which is both time and position dependent. 
One can show that the non-spherical perturbation \eqref{eq:inc_def_disp} is isochoric \cite{gaudron2020jmps}.

The perturbation is governed by the momentum balance equation, expressed in the current, deformed configuration as %Applying the momentum equation based on the current, deformed configuration, we obtain:
\begin{equation}
%\text{div} \boldsymbol{\sigma} = \rho \mathbf{u}_{,tt} \label{eq:mom}
\nabla\cdot \boldsymbol{\sigma} = \rho \mathbf{a} \label{eq:mom}
\end{equation}
where $\nabla\cdot(\bullet)$ is the spatial divergence operator, ${\bf a}$ is the acceleration vector, and $\boldsymbol{\sigma}$ is the Cauchy stress tensor. The traction boundary condition at the bubble wall is 
\begin{equation}
\boldsymbol{\sigma} \mathbf{n}|_{r=R+aY_l^m} = -p_{\rm b}\mathbf{n} - \gamma (\nabla_{\mathcal{S}} \cdot \mathbf{n})\mathbf{n}, \label{eq:bc}
%    \mathbf{n} \cdot \boldsymbol{\sigma} \mathbf{n}|_{(r=R+a Y_l^m)} = p_b + \gamma \nabla \cdot \mathbf{n} \label{eq:bc} % \frac{2 \gamma}{R} - \frac{(l+2)(l-1)}{R^2} a Y_l^m  
\end{equation}
where $[\mathbf{n}] = [-1, a {Y_{l,\theta}^{m}}/R, a Y_{l,\phi}^{m}/(R \text{sin}\theta)]^{\top}$ is the linearized outward unit normal vector on the perturbed bubble wall, and $\nabla_{\mathcal{S}}\cdot(\bullet)$ is the surface divergence operator in the deformed configuration. In the far-field, the stress approaches a state of hydrostatic pressure: $\boldsymbol{\sigma}|_{r\rightarrow\infty} = -p_\infty {\bf I}$, where ${\bf I}$ is the identity tensor.

After inserting \eqref{eq:inc_def_disp} into \eqref{eq:mom}, integrating over $r$ from the current, perturbed bubble wall to the far-field, applying the  radial boundary conditions, and collecting the $O(a)$ terms,  we obtain the governing, second-order differential equation for the bubble instability perturbation magnitude, $a$:
\begin{equation}
a_{,tt} + \eta a_{,t}  - \xi a = 0\,; \quad \eta = \frac{3 R_{,t}}{R} + \frac{4 \mu}{ \rho R^2} + \frac{ l(l+1) \mu}{ 3 \rho R^2}, \label{eq:a_sec_ode}
\end{equation}
where $\xi$  explicitly accounts for  inertial effects during cavitation,  nonlinear deformations of the viscoelastic solid, and  surface tension effects: 
\begin{equation}
\small
    \begin{aligned}
    \xi = & - \frac{R_{,tt}}{R} + \frac{4 \mu R_{,t}}{\rho R^3} -
    \frac{ 2 l(l+1) \mu R_{,t}}{ 3 \rho R^3}  \\
    & - \frac{2 G R_{0} }{\rho R^3} \Big( 1 + \frac{R_{0}^3}{R^3} \Big) - \frac{G l (l+1)}{\rho (R^2 + R R_{0} + R_{0}^2)} \\
    & - \frac{2 \alpha G}{\rho R^2} \frac{(R-R_{0})^2}{R R_{0}} \Big( 1+ \frac{R_{0}}{R} \Big)^3 \Big( 2-  \frac{2R_{0}}{R} +  \frac{3R_{0}^2}{R^2} - \frac{R_{0}^3}{R^3} + \frac{R_{0}^4}{R^4} \Big) \\
    & - \frac{\alpha G l(l+1) (R - R_{0})^2}{5 \rho R R_{0} (R^2 + R R_{0} + R_{0}^2)} \Big( 10 +  \frac{6R_{0}}{R} +  \frac{3R_{0}^2}{R^2} + \frac{R_{0}^3}{R^3} \Big) \\
    & - \frac{(l+2)(l-1)\gamma}{\rho R^3}.
    \end{aligned} \label{eq:xi}
\end{equation}
We note that when $\alpha$\,$ \rightarrow$\,$ 0$, Eqs.\,(\ref{eq:a_sec_ode}-\ref{eq:xi}) describe the evolution of the perturbation magnitude $a$ in a neo-Hookean viscoelastic medium. 
Examining the differential relation \eqref{eq:a_sec_ode}, we find that perturbations $a$ grow  if  $\eta$\,$<$\,$0$ or $\xi$\,$>$\,$0$ and decay if $\eta$\,$>$\,$0$ and $\xi$\,$<$\,$0$ \footnote{See Supplementary Material S5 for more information regarding the stability phase diagram for \eqref{eq:a_sec_ode}}.

%%%%%%%%%%%%%%%%%%%%%%%%%%%%%%%%%%%%%%%%%%%%%%
% Please ignore the format for this figure now. 
% I will re-organize its position at the end before our submission
%\onecolumngrid
%\begin{widetext}
%%%%%%%%%%%%%%%%%%%%%%%%%%%%%%%%%%%%%%%%%%%%%%
\begin{figure} [!t]
\begin{center} 
\includegraphics[width= 0.42 \textwidth]{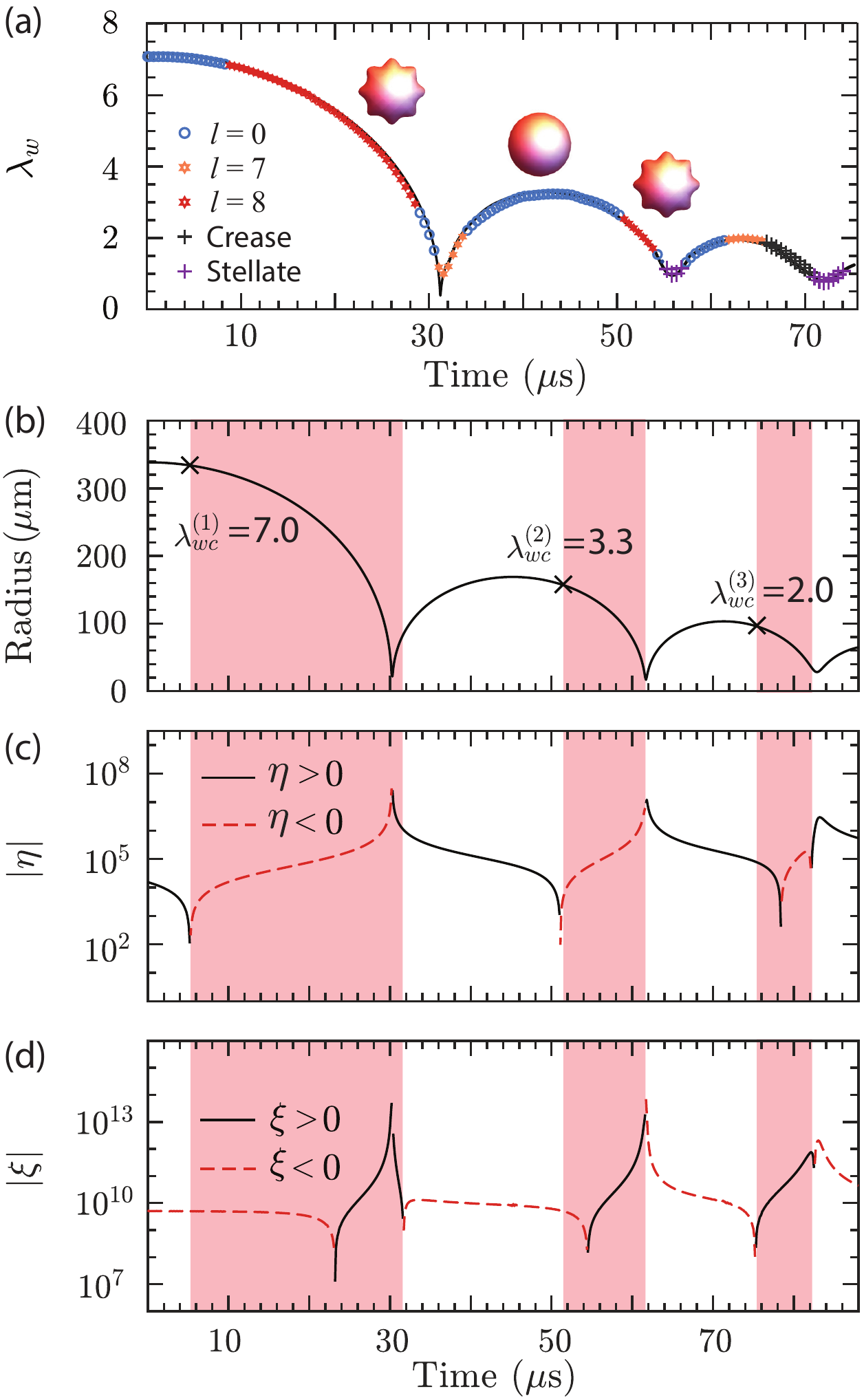}
\end{center}
\caption{(a)\,Hoop stretch at the bubble wall $\lambda_{w}$ for the experiment labeled  ``exp 4'' in Fig.\,\ref{fig:expAll}. Histories of (b)\,the bubble radius $R$ and the quantities (c)\,$\eta$ and (d)\,$\xi$  are numerically computed using simulation parameters based on Fig.\,\ref{fig:expAll} ``exp 4''. Red shaded regions denote situations in which non-spherical instabilities corresponding to $l=8$ are predicted. The critical values of $\lambda_w$ at the onset of  instability occurring during the first three collapse cycles are marked as black crosses in (b).} \label{fig:phase_diagram}
\end{figure}
%%%%%%%%%%%%%%%%%%%%%%%%%%%%%%%%%%%%%%%%%%%%%%
%\twocolumngrid
%\end{widetext}
%%%%%%%%%%%%%%%%%%%%%%%%%%%%%%%%%%%%%%%%%%%%%%
%where $\kappa = l(l+1)$, and the radial stretch ratio at the bubble wall is denoted as $\lambda_{w}={R}/{R_0}$.

Utilizing the radius vs.\ time data from the experimental observations in Fig.\,\ref{fig:expAll} in our instability model Eqs.\,(\ref{eq:a_sec_ode}-\ref{eq:xi}), we can theoretically predict the onset  of various non-spherical instability patterns during the cavitation process. For illustration, based on  this data set (specifically, ``exp 4'' in Fig.~\ref{fig:expAll}), we consider the emergence of instability mode shape $l=8$. The evolution of the bubble radius $R$ for this case is numerically simulated, and the histories of the bubble radius and the quantities $\eta$ and $\xi$ are plotted in Fig.\,\ref{fig:phase_diagram}(b-d). Conditions under which non-spherical instabilities corresponding to $l=8$ are predicted (i.e., $\eta<0$ or $\xi>0$) are marked as red shaded regions in Fig.\,\ref{fig:phase_diagram}(b-d). The onset of non-spherical deformation predicted by the theoretical instability model is in good agreement with the experimental observations in Fig.\,\ref{fig:phase_diagram}(a) during each of the first three collapse cycles.%For this particular data set, our model predicts the emergence and growth of instability shape mode 8, which is highlighted in the red shaded areas in Fig.\,\ref{fig:phase_diagram}(b-d), and is in good agreement with the experimental observations in Fig.\,\ref{fig:exp4}(b-c) and Fig.\,\ref{fig:expAll}. 
%By further computing the phase space variable $\Delta$\,$:=$\,$\eta^2$\,$+$\,$ 4 \xi$ (see Supplementary Material Fig.\,S3(c)), we find that the spatiotemporal evolution of this particular mode, i.e., mode 8, shows moderate growth in time, characterized by an \textit{unstable spiral} type instability first and then a \textit{saddle point} instability in the phase diagram.  

%%%%%%%%%%%%%%%%%%%%%%%%%%%%%%%%%%%%%%%%%%%%%%
%\section{Discussions and Conclusions}\label{sec:results}
%%%%%%%%%%%%%%%%%%%%%%%%%%%%%%%%%%%%%%%%%%%%%%
From Eqs.\,(\ref{eq:a_sec_ode}-\ref{eq:xi}), we find that both material viscosity and surface tension always act to stabilize the bubble against non-spherical deformation. However, when the bubble size is much greater than the characteristic length  $\gamma/G$, the effect of surface tension  is negligible.  When the bubble approaches the final equilibrium radius ($R$\,$\rightarrow$\,$R_{0}$, $R_{,t}$\,$\rightarrow$\,$0$, and $R_{,tt}$\,$\rightarrow$\,$0$), we find that $\eta_{\infty}$\,$>$\,$0$ and  $\xi_{\infty}$\,$<$\,$0$, so that a spherical bubble will be stable under all perturbation modes. Under these conditions, we can also obtain the natural frequency of vibration $\omega_l$ for a bubble in a viscoelastic material corresponding to each non-spherical mode shape $l$:
\begin{equation}
    \omega_l^2 = \frac{4 G}{\rho R_{0}^2} + \frac{G l (l+1)}{3 \rho R_{0}^2} + \frac{(l+2)(l-1) \gamma}{\rho R_{0}^3}.
\end{equation}
As discussed in \cite{murakami2020us}, a non-spherical mode becomes unstable under continuous external driving when the driving frequency $\omega_{d}$ equals $ 2 \omega_{l}$.

%Since our model explicitly tracks all inertial ($\xi^i$), viscoelastic deformation ($\xi^{ve}$), and surface tension contributions from the bubble wall and the surrounding gel, respectively, the ratio between these two contributions, i.e., { $|\xi^{ve}/\xi^{i}|$}, provides a measure of the inertia or solid-deformation dominated regimes during cavitation. Consistent with notions of a violent collapse, inertia dominates the first collapse ({ $|\xi^{ve}/\xi^{i}|$}$ \sim $\,$O(10^{-1})$), whereas  nonlinear viscoelastic material deformations dominate the onset and evolution of instabilities during the remaining subsequent bubble expansion-collapse cycles ($|\xi^{ve}/\xi^{i}|$ $\sim $\,$O(10^{1})$). 
%Similarly, with violent cavitation in water \cite{brenner1995prl}, we have observed the RT instability near the first collapse as shown in leftmost red shaded area in Fig.\,\ref{fig:phase_diagram}(b-d) where inertial effects are dominant and a new non-dimensional ratio $\chi$\,$:=$\,{ $|\xi^{ve}/\xi^{i}|$} takes the value less than $O(10^{-1})$ \footnote{Variables  $\xi^{ve}$,\,$\xi^{i}$, and $\chi$ computed from data set Fig.\,\ref{fig:exp4} are plotted in Supplementary Material Fig.\,S2(a-c).}. 

\begin{figure}[!t]
\begin{center} 
\includegraphics[width=.42 \textwidth]{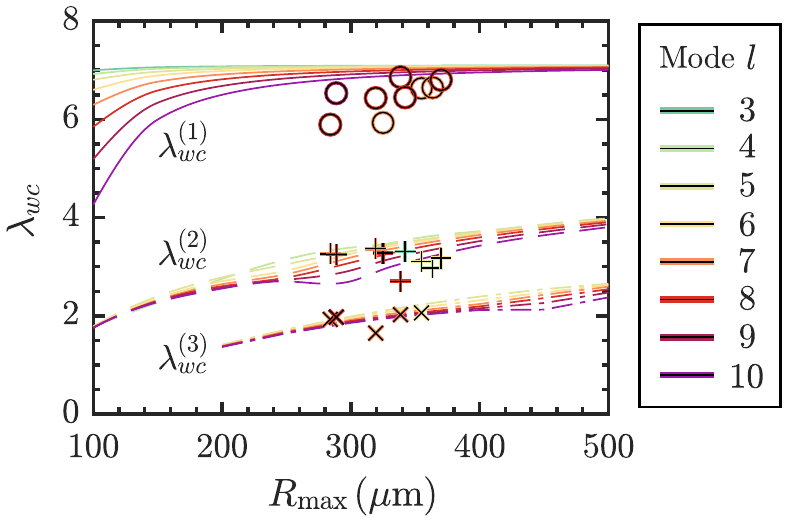}
\end{center}
\caption{Theoretically predicted and experimentally measured values of the critical hoop stress at the bubble wall $\lambda_{wc}$ at the onset of instability during the first three collapse cycles as a function of the  maximum bubble radius $R_\text{max}$ and the non-spherical mode shape  number $l$, where  $\lambda_{\text{max}}$ is fixed at the same value as in Fig.\,\ref{fig:expAll}.} \label{fig:crit_lambda}
\end{figure}

Taking a closer look at the temporal evolution of the experimentally measured bubble wall hoop stretch, $\lambda_w$ (Fig.\,\ref{fig:phase_diagram}(a)), we find that the onset of instabilities occurs for $\lambda_w$\,$>$\,$1$. This marks a significant departure from previous surface instability investigations under quasi-static loading conditions, where rugae patterns require a stretch ratio within the plane of the surface that is less than one \cite{cai2010softmatter,jin2011epl,li2012softmatter,diab2013prsa,2017zhaojap}. 
Intrigued by these observations, we ask how well our instability model predicts the emergence of instabilities for $\lambda_w$\,$ >$\,$ 1$. In Fig.\,\ref{fig:crit_lambda}, we plot the theoretically predicted values of $\lambda_w$ at the onset of instability during the first three collapse cycles as a function of the maximum bubble radius $R_{\text{max}}$ and the non-spherical mode shape number $l$ \footnote{The initial hoop stretch ratio  $\lambda_{\text{max}}$ is fixed at the same value as in Fig.\,\ref{fig:expAll}. Details of the theoretical predictions are summarized in Supplementary Material S7.}.
Next, using the experimentally measured values of $R_{\text{max}}$, we plot the critical values of $\lambda_{w}$ at the onset of each experimentally observed instability occurring in Fig.\,\ref{fig:expAll}, and we find that the critical  values  during the first three collapse cycles are $\lambda_{wc}^{(1)}$\,$=$\,6.5\,$\pm$\,0.3 (circles in Fig.\,\ref{fig:crit_lambda}), $\lambda_{wc}^{(2)}$\,$=$\,3.2\,$\pm$\,0.2 (pluses in  Fig.\,\ref{fig:crit_lambda}), and $\lambda_{wc}^{(3)}$\,$=$\,1.8\,$\pm$\,0.2 (crosses in Fig.\,\ref{fig:crit_lambda}), respectively. Comparing the experimentally obtained values of $\lambda_{wc}$ with the ones predicted by our instability model, we see good  agreement across all  mode shapes, as shown in Fig.\,\ref{fig:crit_lambda}.

%%%%%%%%%%%%%%%%%%%%%%%%%%%%%%%%%%%%%%%%%%%%%%
%%%%%%%%%%%%%%%%%%%%%%%%%%%%%%%%%%%%%%%%%%%%%%

% ----- creases and stellate instabilities -----
It is important to note that other types of non-spherical instabilities are also possible during an inertial cavitation event. For example, in Fig.\,\ref{fig:exp4}(c-vii), one can see crease-like instabilities near the bubble wall for $\lambda_{w}$\,$<$\,1 \cite{milner2017softmatter,bruning2019prl,yang2020semannual}. 
Axisymmetric instabilities may also appear during inertial cavitation in heterogeneous media or in the vicinity of impedance mismatched boundaries \cite{brujan_nahen_schmidt_vogel_2001,bremond2006pof,yang2021semannual_interface}. 
The critical conditions for predicting the onset of these and other non-spherical instabilities remain an active area of research and will be the subject of future work.

Finally, through fully 3D finite element simulations, we find that non-spherical instabilities near the bubble wall can  induce strain and stress amplification in soft materials, which might lead to a thin damage layer developing near the bubble wall \footnote{Fully 3D finite element simulations are presented in Supplementary Material S8.}.
It is also conceivable that within such a layer the material could experience inelastic deformation, fracture, or  significant strain softening \cite{hutchens2016softmatter,movahed2016jasa,cohen_frac_cav}. %\footnote{$I_1$ has been directly measured from experiments, see Supplementary Material S5.}. 
While addressing  appropriate material damage models is an exciting research area beyond the scope presented here, we nevertheless hope that the results of our theory motivate future studies aimed at resolving the intricate mechanics and physics near the bubble wall during these high strain-rate, inertially dominated deformations.

In sum, this paper presents a new theoretical framework for predicting the dynamic onset and evolution of complex non-spherical instability shapes in nonlinear viscoelastic soft materials during inertial cavitation, and provides a new foundation for characterizing and classifying dynamic instabilities under extreme  loading conditions.

%%%%%%%%%%%%%%%%%%%%%%%%%%%%%%%%%%%%%%%%%%%%%%%%%%%%%%%%%
%\section{Acknowledgment}\label{sec:ack}
We gratefully acknowledge funding support from the Office of Naval Research (Dr. Timothy Bentley) under grant N000141712058. We thank Dr. Lauren Hazlett for helpful discussions and editing of the manuscript.
 
%\section*{References}
%\bibliographystyle{aipauth4-1}
\bibliography{reference}  

\end{document}

\title{SUPPLEMENTARY MATERIAL of \\
``Predicting complex Instability Shapes of \\
Inertial Cavitation Bubbles in Viscoelastic Soft Matter"}% 

\author{Jin Yang}
\affiliation{Department of Mechanical Engineering, University of Wisconsin-Madison}

\author{Anastasia Tzoumaka}%
\affiliation{School of Engineering, Brown University}

\author{Kazuya Murakami}%
\affiliation{Department of Mechanical Engineering, University of Michigan, Ann Arbor}

\author{Eric Johnsen}%
\affiliation{Department of Mechanical Engineering, University of Michigan, Ann Arbor}

\author{David L. Henann}%
\affiliation{School of Engineering, Brown University}

\author{Christian Franck} \email{cfranck@wisc.edu; Corresponding author}
\affiliation{Department of Mechanical Engineering, University of Wisconsin-Madison}

\date{\today} 
\maketitle
\onecolumngrid

%%%%%%%%%%%%%%%%%%%%%%%%%%%%%%%%%%%%%%%%%%%%%%%%%%%%%%%
\section{Polyacrylamide Hydrogel Preparation}
\label{suppsec:PA_protocol}
To prepare polyacrylamide (PA) hydrogels, 40.0 \% acrylamide solution and 2.0 \% bis solution (Bio-Rad, Hercules, CA) are mixed to a final concentration of 8.0 \%/0.08 \% Acrylamide/Bis (v/v) in deionized water and crosslinked with 0.5 \% Ammonium Persulfate (APS) and 1.25 \% N,N,N',N'-tetramethylethane-1,2-diamine (TEMED)(ThermoFisher Scientific, USA).  Once mixed, PA samples were allowed to
polymerize in custom 15 mm diameter by 2 mm height cylindrical delrin molds for 1 hour. PA hydrogels were then submerged in deionized water 24 hours before testing to allow for any swelling to equilibrate.
The viscoelastic material properties of the final prepared PA hydrogels are characterized by a quadratic law strain stiffening Kelvin-Voigt (qKV) viscoelastic model with ground-state shear modulus $G=2.77$ kPa, viscosity $\mu=0.186 \pm 0.194 $ Pa$\cdot$s, and strain stiffening parameter $\alpha=0.48 \pm 0.14$ \cite{yang2020eml}.

%%%%%%%%%%%%%%%%%%%%%%%%%%%%%%%%%%%%%%%%%%%%%%%%%%%%%%
\section{Magnitude of bubble surface perturbations}
\begin{figure}[!t]
\centering
\includegraphics[width = 0.9 \textwidth]{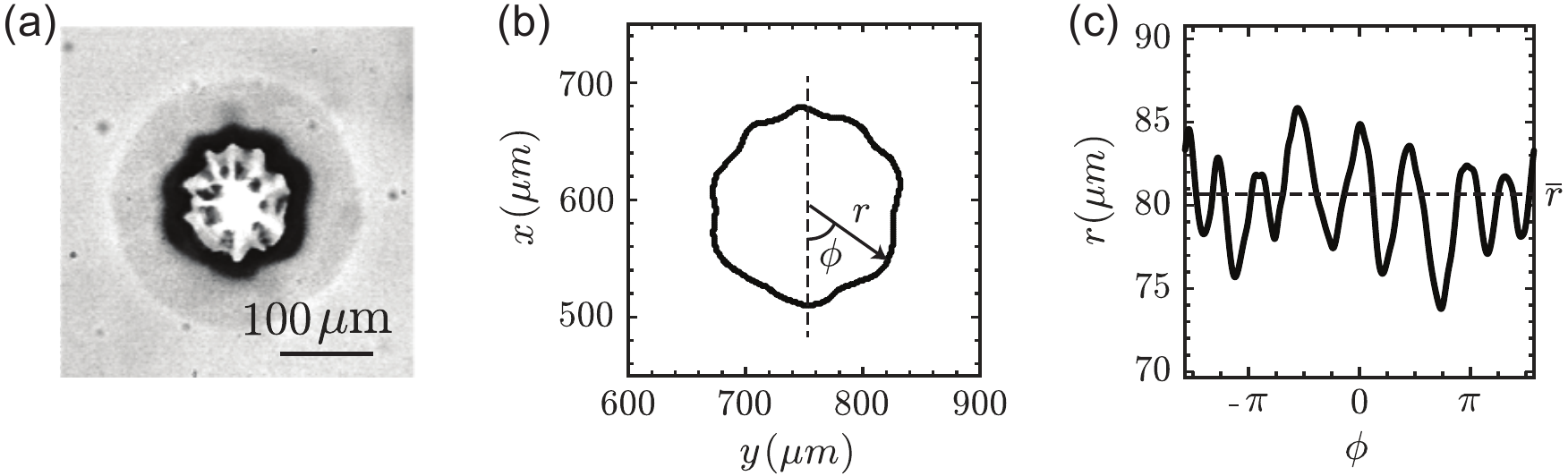}
\caption{Measurement of magnitude of bubble shape instabilities. (a)\,A typical frame from the recorded experimental video. (b)\,Bubble shape is extracted from the image post-processing of each frame. (c)\,Bubble surface is further transformed into a polar coordinate system.} \label{fig:mag_wrk}
\end{figure}

This section describes how we quantitatively measure the magnitude of bubble surface perturbations in our experiments. First, we perform image post-processing to extract the outline of bubble wall from each video frame, see Fig.\,\ref{fig:mag_wrk}(a-b), which is then rewritten into  a polar coordinate system $\lbrace r, \phi \rbrace$, see Fig.\,\ref{fig:mag_wrk}(c), where $r$ and $\phi$ are the radial and polar angular coordinates and defined in Fig.\,\ref{fig:mag_wrk}(b). We further define the mean radius of the bubble as:
\begin{equation}
    \bar{r} =  \frac{1}{2 \pi} \int_{\phi} r d \phi
\end{equation}

The upper bound, lower bound, mean, and root-mean-square (RMS) values of the magnitude of the bubble surface perturbation are further defined as:
\begin{align}
    a_{\text{upper bound}} &= \max_{\phi} |r-\bar{r}|, \\
    a_{\text{lower bound}} &= \min_{\phi} |r-\bar{r}|, \\
    a_{\text{mean}} &= \frac{1}{2 \pi} \int_{\phi} |r-\bar{r}| d \phi, \\
    a_{\text{RMS}} &= \left( \frac{1}{2 \pi} \int_{\phi} |r-\bar{r}|^2 d \phi \right)^{1/2}.
\end{align}

%%%%%%%%%%%%%%%%%%%%%%%%%%%%%%%%%%%%%%%%%%%%%%%%%%%%%%
%%%%%%%%%%%%%%%%%%%%%%%%%%%%%%%%%%%%%%%%%%%%%%%%%%%%%% 

%%%%%%%%%%%%%%%%%%%%%%%%%%%%%%%%%%%%%%%%%%%%%%%%%%%%%%
\section{Derivation of the ansatz for the perturbed displacement field in main text Eq (4)}

\begin{figure}[!b]
\centering
\includegraphics[width = 0.27 \textwidth]{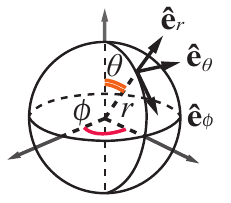}
\caption{Spherical coordinates $\lbrace r, \theta, \phi \rbrace$. } \label{fig:sph_coord}
\end{figure}

Consider a spherical void (see Fig.\,\ref{fig:sph_coord}) with reference configuration $\mathcal{B}_0(r_0,\theta_0,\phi_0)$, 
$\lbrace R_{0} \leqslant r_0 \leqslant \infty, \ 0  \leqslant \theta_0 \leqslant \pi, \  0  \leqslant \phi_0 \leqslant 2 \pi \rbrace$, and current deformed configuration $\mathcal{B}(r,\theta,\phi)$,  $\lbrace R   \leqslant r \leqslant \infty, \ 0  \leqslant \theta \leqslant \pi, \ 0  \leqslant \phi \leqslant 2 \pi \rbrace$, where $\lbrace r_0, r \rbrace$ represent referential and current radial coordinates, $\lbrace \phi_0, \phi \rbrace$ are referential and current azimuthal angular coordinates, and $\lbrace \theta_0, \theta \rbrace$ are referential and current polar angular coordinates.
The time-dependent bubble radius is $R(t)$, and $R_0$ denotes the undeformed bubble radius.
%
%%%%%%%%%%%%%%%%%%%%%%%%%%%%%%%%%%%%%%%%%%%%%%%%%%%%%%
%%%%%%%%%%%%%%%%%%%%%%%%%%%%%%%%%%%%%%%%%%%%%%%%%%%%%%
%
The spherically symmetric base state of deformation is described by the radial mapping $r=(r_0^3 + R^3(t) - R_0^3)^{1/3}$, so that in the base state, the only non-zero component of displacement is $u_{r} =r- r_{0}$. We then consider a displacement perturbation of the form $\{\tilde{u}_r,\tilde{u}_\theta,\tilde{u}_\phi\}$, which is added to the base-state displacement field. The resulting deformation gradient is ${\bf F} = {\bf F}_0+{\bf d}$, where ${\bf F}_0$ is the base-state deformation gradient, which can be expressed referentially as
\begin{equation}
[\mathbf{F}_0] = \begin{bmatrix}
\frac{r_0^2}{(r_0^3 + R^3 - R_0^3)^{2/3}} & 0 & 0 \\[4pt]
0 & \frac{(r_0^3+R^3-R_0^3)^{1/3}}{r_0} & 0 \\[4pt]
0 & 0 & \frac{(r_0^3+R^3-R_0^3)^{1/3}}{r_0}
\end{bmatrix}\quad\text{so that}\quad \det{\bf F}_0=1, \label{eq:F} 
\end{equation} 
and ${\bf d}$ is the perturbation of the deformation gradient, which can be expressed spatially as
\begin{equation}
[\mathbf{d}] = \begin{bmatrix}
\frac{\partial \tilde{u}_r}{\partial r} & \frac{1}{r} \frac{\partial \tilde{u}_r}{\partial \theta} - \frac{\tilde{u}_{\theta}}{r} & \frac{1}{r \text{sin}\theta}  \frac{\partial \tilde{u}_r}{\partial \phi} - \frac{\tilde{u}_{\phi}}{r} \\[4pt]
\frac{\partial \tilde{u}_{\theta}}{\partial r} & \frac{1}{r} \frac{\partial \tilde{u}_{\theta}}{\partial \theta} + \frac{\tilde{u}_{r}}{r} & \frac{1}{r \text{sin}\theta}  \frac{\partial \tilde{u}_{\theta}}{\partial \phi} - \frac{\tilde{u}_{\phi} \text{cot}\theta}{r} \\[4pt]
\frac{\partial \tilde{u}_{\phi}}{\partial r} & \frac{1}{r} \frac{\partial \tilde{u}_{\phi}}{\partial \theta} & \frac{1}{r \text{sin}\theta}  \frac{\partial \tilde{u}_{\phi}}{\partial \phi} + \frac{\tilde{u}_r}{r} + \frac{\tilde{u}_{\theta} \text{cot}\theta}{r}
\end{bmatrix}. \label{eq:inc_d}
\end{equation}
%
%In the undisturbed base state where spherical symmetry holds, the deformation gradient $\mathbf{F}$ and its increment $\mathbf{d}$ have the following explicit forms in spherical coordinates \cite{Ogden2007}:
%\begin{equation}
%\mathbf{F} = \begin{bmatrix}
%\frac{\partial r}{\partial r_0} & 0 & 0 \\
%0 & \frac{r}{r_0} & 0 \\
%0 & 0 & \frac{r}{r_0}
%\end{bmatrix};  \ J = \text{det}(\mathbf{F}) \label{eq:F} 
%\end{equation} 
%\vspace{-0.2cm}
%\begin{equation}
%\mathbf{d} = \begin{bmatrix}
%\frac{\partial u_r}{\partial r} & \frac{1}{r} \frac{\partial u_r}{\partial \theta} - \frac{u_{\theta}}{r} & \frac{1}{r \text{sin}\theta}  \frac{\partial u_r}{\partial \phi} - \frac{u_{\phi}}{r} \\
%\frac{\partial u_{\theta}}{\partial r} & \frac{1}{r} \frac{\partial u_{\theta}}{\partial \theta} + \frac{u_{r}}{r} & \frac{1}{r \text{sin}\theta}  \frac{\partial u_{\theta}}{\partial \phi} - \frac{u_{\phi} \text{cot}\theta}{r} \\
%\frac{\partial u_{\phi}}{\partial r} & \frac{1}{r} \frac{\partial u_{\phi}}{\partial \theta} & \frac{1}{r \text{sin}\theta}  \frac{\partial u_{\phi}}{\partial \phi} + \frac{u_r}{r} + \frac{u_{\theta} \text{cot}\theta}{r}
%\end{bmatrix} \label{eq:inc_d}
%\end{equation}
%
Since the surrounding material is assumed to be  incompressible, the incremental deformation gradient $\mathbf{d}$ must satisfy
\begin{equation}
\text{tr}(\mathbf{d}) = \frac{\partial \tilde{u}_r}{\partial r} + \frac{1}{r} \frac{\partial \tilde{u}_{\theta}}{\partial \theta} + \frac{\tilde{u}_r}{r} + \frac{1}{r \text{sin} \theta} \frac{\partial \tilde{u}_{\phi}}{\partial \phi} + \frac{\tilde{u}_r}{r} + \frac{\tilde{u}_{\theta} \text{cot} \theta}{r}  = 0  \label{eq:incomp_sph}
\end{equation}

We consider the following displacement perturbation:
\begin{align}
\left\lbrace
\begin{aligned}
\tilde{u}_{r} &= f(r) \text{cos}( m \phi ) P_{l}^{m}(\text{cos} \theta),  \quad  \\
\tilde{u}_{\theta} &= g(r) \text{cos}( m \phi ) \frac{d}{d \theta} P_{l}^{m}(\text{cos} \theta),  \\
\tilde{u}_{\phi} &= h(r) m \text{sin}( m \phi ) P_{l}^{m}(\text{cos} \theta) ,
\end{aligned}
\right. \label{eq:inc_def_all}
\end{align}
where $f(r)$, $g(r)$, and $h(r)$ are functions that only depend on the spatial radial coordinate $r$. 
%
%\begin{align}
%\left\lbrace
%\begin{aligned}
%u_{r} &= f(r) \text{cos}( m \phi ) P_{l}^{m}(\text{cos} \theta)  \quad  \\
%u_{\theta} &= g(r) \text{cos}( m \phi ) \frac{d}{d \theta} P_{l}^{m}(\text{cos} \theta)  \\
%u_{\phi} &= e(r) m \text{sin}( m \phi ) P_{l}^{m}(\text{cos} \theta) \\
%\dot{p} &= h (r) \text{cos}( m \phi ) P_{l}^{m}(\text{cos} \theta)
%\end{aligned}
%\right. \label{eq:inc_def_all}
%\end{align}
%
Perturbations of this form are related to the normalized spherical harmonic functions $Y_{l}^{m}(\theta,\phi) := \text{cos}(m \phi) P_{l}^{m} (\text{cos}\theta) $, where $P_{l}^{m} (\text{cos}\theta)$ are the normalized associated Legendre polynomials, $m$ is the fixed wavenumber in the azimuthal angular coordinate,  and the integers $m$ and $l$ satisfy $|m|  \leqslant l$.

After inserting the ansatz \eqref{eq:inc_def_all} into the incompressibility condition \eqref{eq:incomp_sph},
we obtain
\begin{equation}
\frac{\partial f}{\partial r} + \frac{2f}{r} - \frac{g}{r}\left[l(l+1)-\frac{m^2}{\text{sin}^2 \theta} \right] + \frac{m^2 h}{r \text{sin}\theta}  = 0. \label{eq:inc_def_incomp_cond}
\end{equation}
Here, motivated by our experimental observations, we focus on non-spherical, non-axisymmetric instabilities, in which $m>0$. Thus, the solution of Eq.\,\eqref{eq:inc_def_incomp_cond} obeys following conditions:
%
%Here, we focus on the non-spherical non-axisymmetric instabilities ($m$\,$>$\,$ 0$), which are what we experimentally observed. Thus the solution of (\ref{eq:inc_def_incomp_cond}), independent of variable $\theta$, obey following conditions:
\begin{equation}
g=h=0 \quad \Rightarrow \quad \tilde{u}_{\theta}=\tilde{u}_{\phi} = 0   ,
\end{equation}
\begin{equation}
\frac{\partial f}{\partial r} + \frac{2f}{r}  = 0 \
\Rightarrow \ f(r) = C r^{-2}. \label{eq:case1_incomp}
\end{equation}
We further rewrite $C$ as $a R^2$, where $a(t)$ is the time-dependent perturbation magnitude at the bubble wall, so that \eqref{eq:inc_def_all} takes the form of Eq.\,(4) in the main text. %$a$ is a time-dependent coefficient, but it is independent of the position $r$. Then (\ref{eq:inc_def_all}) can be further simplified as (4) in the main text. 

%%%%%%%%%%%%%%%%%%%%%%%%%%%%%%%%%%%%%%%%%%%%%%%%%%%%%%
\section{Derivation of Eqs (7-8) in the main text}
\subsection{Kinematics}
First, we calculate several kinematical fields in the surrounding soft material.   
For the ansatz for the perturbed displacement field in main text Eq.\,(4), the deformation mapping is 
\begin{equation}
\left\lbrace 
\begin{aligned}
r(r_0,\theta_0,\phi_0,t) &= (r_0^3 + R^3(t) - R_0^3)^{1/3} + \frac{a(t)R^2(t)}{(r_0^3 + R^3(t) - R_0^3)^{2/3}} Y_{l}^{m} (\theta_0,\phi_0), \\
\theta(r_0,\theta_0,\phi_0,t) &= \theta_0, \\
\phi(r_0,\theta_0,\phi_0,t) &= \phi_0 ,
\end{aligned}
\right. \label{eq:def_mapping}
\end{equation}
where $Y_{l}^{m} (\theta,\phi)$
%  := $\text{cos}(m \phi) P_l^m (\text{cos}\theta)$ 
is the normalized $(l,m)^{th}$ spherical harmonic function, %, and $l,m$ are the mode order and degree respectively.  
%$P_l^m (\text{cos}\theta)$ is the associated Legendre polynomial function. 
and $a(t)$ is a time-dependent coefficient, which is assumed to be much smaller than $R(t)$. 
The referential Lagrangian description of the displacement field is then
\begin{equation}
\left\lbrace 
\begin{aligned}
u_r(r_0,\theta_0,\phi_0,t) &= (r_0^3 + R^3(t) - R_0^3)^{1/3} - r_0 + \frac{a(t)R^2(t)}{(r_0^3 + R^3(t) - R_0^3)^{2/3}} Y_{l}^{m} (\theta_0,\phi_0), \\
u_{\theta}(r_0,\theta_0,\phi_0,t) &= 0, \\
u_{\phi}(r_0,\theta_0,\phi_0,t) &= 0 .
\end{aligned}
\right. \label{eq:inc_def_disp}
\end{equation}
The deformation gradient tensor in the referential Lagrangian description can be simplified as
\begin{equation}
\begin{aligned}
[\mathbf{F}] &= \begin{bmatrix}
\frac{\partial r}{\partial r_0} & \frac{1}{r_0}\frac{\partial u_r}{\partial \theta_0} & \frac{1}{r_0 \text{sin} \theta_0} \frac{\partial u_r}{\partial \phi_0} \\[4pt]
0 & \frac{r}{r_0} & 0 \\[4pt]
0 & 0 & \frac{r}{r_0} 
\end{bmatrix}  \\
&= \begin{bmatrix}
      \frac{r_0^2}{(r_0^3+R^3-R_0^3)^{2/3}} - \frac{2 r_0^2 a R^2 Y_l^m}{(r_0^3+R^3-R_0^3)^{5/3}}   &  \frac{a R^2}{r_0 (r_0^3+R^3-R_0^3)^{2/3}} \frac{\partial Y_l^m}{\partial \theta_0}  &  \frac{a R^2}{r_0 \text{sin}\theta_0 (r_0^3+R^3-R_0^3)^{2/3}} \frac{\partial Y_l^m }{\partial \phi_0 } \\[4pt]
      0  & \frac{(r_0^3+R^3-R_0^3)^{1/3}}{r_0} + \frac{a R^2 Y_{l}^{m}}{r_0 (r_0^3+R^3-R_0^3)^{2/3}}  &  0 \\[4pt]
      0  &  0 &  \frac{(r_0^3+R^3-R_0^3)^{1/3}}{r_0} + \frac{a R^2 Y_{l}^{m}}{r_0 (r_0^3+R^3-R_0^3)^{2/3}}
    \end{bmatrix}.
\end{aligned}
\end{equation}
Since $a(t) \ll R(t)$, the left Cauchy-Green deformation tensor can be linearized as
\begin{equation}
    \begin{aligned}
    [\mathbf{B}] &= [\mathbf{F}][\mathbf{F}]^{\top} \\
    &\approx \begin{bmatrix}
     \frac{r_0^4}{(r_0^3+R^3-R_0^3)^{4/3}} - \frac{4 r_0^4 a R^2 Y_l^m}{(r_0^3+R^3-R_0^3)^{7/3}}   &  \frac{a R^2}{r_0^2 (r_0^3+R^3-R_0^3)^{1/3}} \frac{\partial Y_l^m}{\partial \theta_0}  &  \frac{a R^2}{r_0^2 \text{sin}\theta_0 (r_0^3+R^3-R_0^3)^{1/3}} \frac{\partial Y_l^m }{\partial \phi_0 } \\[4pt]
      \frac{a R^2 }{r_0^2 (r_0^3+R^3-R_0^3)^{1/3} }\frac{\partial Y_l^m}{\partial \theta_0}  & \frac{(r_0^3+R^3-R_0^3)^{2/3}}{r_0^2} + \frac{2a R^2 Y_{l}^{m}}{r_0^2 (r_0^3+R^3-R_0^3)^{1/3}}  &  0 \\[4pt]
      \frac{a R^2}{r_0^2 \text{sin}\theta_0 (r_0^3+R^3-R_0^3)^{1/3}} \frac{\partial Y_l^m }{\partial \phi_0 }  &  0 &  \frac{(r_0^3+R^3-R_0^3)^{2/3}}{r_0^2} + \frac{2a R^2 Y_{l}^{m}}{r_0^2 (r_0^3+R^3-R_0^3)^{1/3}}
    \end{bmatrix}.
    \end{aligned}
\end{equation}
To express kinematical fields using a spatial description, we linearize the inverse deformation mapping:
\begin{equation}
    r_0(r,\theta,\phi,t) \approx (r^3+R_0^3-R^3(t))^{1/3} - \frac{a(t) R^2(t)}{(r^3+R_0^3-R^3(t))^{2/3}} Y_l^m(\theta, \phi).
\end{equation}
Then, using the inverse mapping and linearizing, the left Cauchy-Green deformation tensor can be expressed spatially as
\begin{equation}\label{eq:spatialB}
    \begin{split}
        [\mathbf{B}] \approx \begin{bmatrix}
     \frac{(r^3+R_0^3-R^3)^{4/3}}{r^4}-\frac{4 a R^2 (r^3+R_0^3-R^3)^{1/3} Y_l^m}{r^4}   &  \frac{a R^2}{r (r^3+R_0^3-R^3)^{2/3}} \frac{\partial Y_l^m}{\partial \theta} & \frac{a R^2}{r \text{sin}\theta (r^3+R_0^3-R^3)^{2/3}} \frac{\partial Y_l^m}{\partial \phi} \\[4pt]
     \frac{a R^2}{r (r^3+R_0^3-R^3)^{2/3}} \frac{\partial Y_l^m}{\partial \theta}  & \frac{r^2}{(r^3+R_0^3-R^3)^{2/3}} + \frac{2 a R^2 r^2 Y_l^m}{(r^3+R_0^3-R^3)^{5/3}} & 0 \\[4pt]
      \frac{a R^2}{r \text{sin}\theta (r^3+R_0^3-R^3)^{2/3}} \frac{\partial Y_l^m}{\partial \phi}  & 0 & \frac{r^2}{(r^3+R_0^3-R^3)^{2/3}} + \frac{2 a R^2 r^2 Y_l^m}{(r^3+R_0^3-R^3)^{5/3}}
        \end{bmatrix}.
    \end{split}
\end{equation}

%For an incompressible material, we can further define radial stretch ratio, $\lambda$, as
%\begin{equation}
%\begin{split}
%\lambda &= \frac{r}{r_0} = \frac{\left( r_0^3 + R^3 - R_{0}^3 \right)^{\frac{1}{3}}}{r_0} \\
%&\approx \frac{r}{(r^3+R_0^3-R^3)^{1/3}} - \frac{a R^2 Y_l^m}{r^2 (r^3+R_0^3-R^3)^{1/3}} + \frac{a R^2 r Y_l^m}{(r^3+R_0^3-R^3)^{4/3}}
%\end{split}
%\end{equation}

The referential description of the radial velocity is
\begin{equation}
    V_r(r_0,\theta_0,\phi_0,t) = \frac{\partial u_r}{\partial t}\bigg\rvert_{r_0} = \frac{R^2 R_{,t}}{(r_0^3+R^3-R_0^3)^{2/3}} - \frac{2a R^4 R_{,t} Y_l^m}{(r_0^3+R^3-R_0^3)^{5/3}} + \frac{(a_{,t}R^2 + 2 a R R_{,t}) Y_l^m}{(r_0^3+R^3-R_0^3)^{2/3}},
\end{equation}
and upon linearization, the spatial description of the radial velocity is 
\begin{equation}
    \begin{aligned}
    v_r(r,\theta,\phi,t) \approx \frac{R^2 R_{,t}}{r^2} + \frac{a_{,t}R^2 + 2a R R_{,t}}{r^2} Y_l^m.
    \end{aligned}
\end{equation}
The spatial velocity gradient tensor is 
\begin{equation}
    [\mathbf{L}] = [\nabla \mathbf{v}] = \begin{bmatrix}
    -\frac{2 R^2 R_{,t}}{r^3} - \frac{2 (a_{,t}R^2 + 2 a R R_{,t})}{r^3} Y_l^m & \frac{(a_{,t}R^2 + 2 a R R_{,t})}{r^3} \frac{\partial Y_l^m}{\partial \theta} & \frac{(a_{,t}R^2 + 2 a R R_{,t})}{r^3 \text{sin}\theta} \frac{\partial Y_l^m}{\partial \phi} \\[4pt]
    0 & \frac{R^2 R_{,t}}{r^3} + \frac{(a_{,t}R^2 + 2 a R R_{,t})}{r^3} Y_l^m & 0 \\[4pt]
    0 & 0 & \frac{R^2 R_{,t}}{r^3} + \frac{(a_{,t}R^2 + 2 a R R_{,t})}{r^3} Y_l^m
    \end{bmatrix},
\end{equation}
and the rate of deformation tensor $\mathbf{D}$ is then 
\begin{equation}\label{eq:spatialD}
\begin{aligned}
    [\mathbf{D}] &= \frac{1}{2}\left( [\mathbf{L}] + [\mathbf{L}]^{\top} \right) \\
    &= \begin{bmatrix}
    -\frac{2 R^2 R_{,t}}{r^3}-\frac{2 (a_{,t}R^2+2a R R_{,t})}{r^3} Y_l^m & \frac{(a_{,t}R^2+2a R R_{,t})}{2 r^3} \frac{\partial Y_l^m}{\partial \theta} &  \frac{(a_{,t}R^2+2a R R_{,t})}{2 r^3 \text{sin}\theta} \frac{\partial Y_l^m}{\partial \phi} \\[4pt]
    \frac{(a_{,t}R^2+2a R R_{,t})}{2 r^3} \frac{\partial Y_l^m}{\partial \theta} & \frac{ R^2 R_{,t}}{r^3}+\frac{ (a_{,t}R^2+2a R R_{,t})}{r^3} Y_l^m & 0 \\[4pt]
    \frac{(a_{,t}R^2+2a R R_{,t})}{2 r^3 \text{sin}\theta} \frac{\partial Y_l^m}{\partial \phi} & 0 & \frac{ R^2 R_{,t}}{r^3}+\frac{ (a_{,t}R^2+2a R R_{,t})}{r^3} Y_l^m 
    \end{bmatrix}.
    \end{aligned}
\end{equation}
The referential description of the radial acceleration is 
\begin{equation}
\begin{aligned}
    A_r(r_0,\theta_0,\phi_0,t) = \frac{\partial v_r}{\partial t}\bigg\rvert_{r_0} &=  \frac{2R R_{,t}^2 + R^2 R_{,tt}}{(r_0^3+R^3-R_0^3)^{2/3}} -  \frac{2 R^4 R_{,t}^2 }{ (r_0^3+R^3-R_0^3)^{5/3}}  \\
%    & \quad - \frac{2a(4 R^3 R_{,t}^2 + R^4 R_{,tt}) + 2 a_{,t} R^4 R_{,t}}{(r_0^3+R^3-R_0^3)^{5/3}} Y_l^m - \frac{2 (a_{,t}R^2 + 2 a R R_{,t}) R^2 R_{,t}}{(r_0^3+R^3-R_0^3)^{5/3}} Y_l^m \\
    & \quad + \frac{(a_{,tt}R^2 + 4 a_{,t} R R_{,t} + 2 a R_{,t}^2 + 2 a R R_{,tt})}{(r_0^3+R^3-R_0^3)^{2/3}}Y_l^m\\
    & \quad - \frac{2(2 a_{,t} R^4 R_{,t} + 6a R^3 R_{,t}^2 + aR^4 R_{,tt})}{(r_0^3+R^3-R_0^3)^{5/3}} Y_l^m + \frac{10 a R^6 R_{,t}^2 }{(r_0^3+R^3-R_0^3)^{8/3}}Y_l^m,
    \end{aligned}
\end{equation}
and upon linearization, the spatial description of the radial acceleration is 
\begin{equation}
    \begin{aligned}
    a_{r}(r,\theta,\phi,t) = \frac{2 R R_{,t}^2+ R^2 R_{,tt}}{r^2} - \frac{2 R^4 R_{,t}^2}{r^5} + \frac{(a_{,tt}R^2 + 4 a_{,t}R R_{,t} + 2 a R_{,t}^2 + 2 a R R_{,tt})}{r^2} Y_l^m - \frac{(4 a_{,t} R^4 R_{,t}+ 8 a R^3 R_{,t}^2)}{r^5} Y_l^m.
    \end{aligned} \label{eq:a_spatial}
\end{equation}

\subsection{Momentum balance equation}
The momentum balance equation is 
\begin{equation}
\nabla \cdot \boldsymbol{\sigma} = \rho \mathbf{a}, \label{eq:bubble_dyn_mom_eq}
\end{equation}
where $\nabla\cdot(\bullet)$ is the spatial divergence operator; $\rho$ is the spatial mass density; ${\bf a}$ is the acceleration vector;  and $ \boldsymbol{\sigma}$ is the Cauchy stress tensor, which has the following explicit form for the quadratic law Kelvin-Voigt (qKV)  material model:
\begin{equation}\label{eq:stress}
     \boldsymbol{\sigma} = G [1+ \alpha (I_1-3)] \mathbf{B} + 2 \mu \mathbf{D} - p \mathbf{I},
\end{equation}
where $p$ is a pressure-like quantity arising due to the incompressibility constraint in the surrounding material, and $I_1$ is the first invariant of the left Cauchy-Green deformation tensor: 
\begin{equation}
    I_1 = \text{tr}(\mathbf{B}) = \frac{(r^3+R_0^3-R^3)^{4/3}}{r^4} + \frac{2 r^2}{(r^3+R_0^3-R^3)^{2/3}} - \frac{4 a R^2 (r^3+R_0^3-R^3)^{1/3}}{r^4} Y_l^m + \frac{4 a R^2 r^2}{(r^3+R_0^3-R^3)^{5/3}} Y_l^m. 
\end{equation}
We note that the pressure-like quantity $p$ consists of the sum of the base-state field, which only depends on $r$, and a perturbation field, which depends on $r$, $\theta$, and $\phi$. In what follows, we focus on the radial component of the momentum balance equation \eqref{eq:bubble_dyn_mom_eq}:
\begin{equation}\label{eq:mombal_radial}
    \frac{\partial \sigma_{rr}}{\partial r} + \dfrac{1}{r}\dfrac{\partial \sigma_{r\theta}}{\partial \theta} + \dfrac{1}{r\sin\theta}\dfrac{\partial\sigma_{r\phi}}{\partial\phi} + \dfrac{2}{r}(\sigma_{rr}-\sigma_{\theta\theta}) + \dfrac{\cot\theta}{r}\sigma_{r\theta} = \rho a_r,
\end{equation}
where we have recognized that $\sigma_{\theta\theta}=\sigma_{\phi\phi}$.

% ======================================================
\subsection{Boundary conditions}
The traction boundary condition at the bubble wall $(r = R+aY_l^m)$ is
\begin{equation}
    \boldsymbol{\sigma} \mathbf{n}|_{r=R+aY_l^m} = -p_{\rm b}\mathbf{n} - \gamma (\nabla_{\mathcal{S}} \cdot \mathbf{n})\mathbf{n} ,  \label{eq:bc_R}  
\end{equation}
where $[\mathbf{n}] = [-1, a {Y_{l,\theta}^{m}}/R, a Y_{l,\phi}^{m}/(R \text{sin}\theta)]^{\top}$ is the linearized outward unit normal vector on  the bubble wall, $p_{\rm b}$ is the pressure inside the bubble, and $\nabla_{\mathcal{S}}\cdot(\bullet)$ is the surface divergence operator in the deformed configuration. After linearization \cite{plesset1954jap}, the radial component of Eq.~\eqref{eq:bc_R} is
\begin{equation}\label{eq:wallBC_radial}
\sigma_{rr}|_{r=R+aY^m_l} = - p_b + \frac{2 \gamma}{R} + \frac{(l+2)(l-1)}{R^2} a Y_l^m.
\end{equation}
In the far-field, the perturbation decays to zero, and the stress approaches a state of hydrostatic pressure: $\boldsymbol{\sigma}|_{r\rightarrow\infty} = -p_\infty {\bf I}$, where $p_{\infty}$ is the far-field atmospheric pressure. Therefore, the far-field boundary condition for the radial stress component is
\begin{equation}
    \sigma_{rr}|_{r \rightarrow \infty} = -p_{\infty}. \label{eq:bc_infty}
\end{equation}
%We also assume that there is a stress free boundary condition as $r$\,$ \rightarrow$\,$ \infty$ since the introduced perturbation decays to zero at the far-field:
%\begin{equation}
%    \mathbf{n} \cdot \boldsymbol{\sigma} \mathbf{n}|_{(r \rightarrow \infty)} = p_{\infty} \label{eq:bc_infty}
%\end{equation}
%where $p_{\infty}$ is the far-field atmospheric pressure.

\subsection{Results}
Using \eqref{eq:spatialB} and \eqref{eq:spatialD}, we insert \eqref{eq:stress} and \eqref{eq:a_spatial} into \eqref{eq:mombal_radial}, integrate over $r$ from the current, perturbed bubble wall $(R+a Y_l^m)$ to the far-field $(r\rightarrow\infty)$ and apply the radial boundary conditions \eqref{eq:wallBC_radial} and \eqref{eq:bc_infty}. After collecting the $O(1)$ terms, we recover the Rayleigh-Plesset governing equation for the dynamics of the unperturbed bubble radius in a nonlinear strain-stiffening viscoelastic material, which is consistent with our previous study as shown in Yang, et al.\,\cite{yang2020eml}:
\begin{equation}\label{eq:RP}
R R_{,tt} + \frac{3}{2} R_{,t}^2 = \frac{1}{\rho} \left( p_b - p_{\infty} + S - \frac{2 \gamma}{R} \right),
\end{equation}
where stress integral $S$ takes the form
\begin{equation}
 S = \frac{(3 \alpha - 1)G}{2}  \left[ 5  - \left( \frac{{R}_{0}}{R} \right)^4  - \frac{4 {R}_{0}}{R} \right] - \frac{4 \mu \dot{R}}{R} +  2 \alpha G  \left[\frac{27}{40} + \frac{1}{8} \left( \frac{{R}_{0}}{R} \right)^8 + \frac{1}{5}\left( \frac{{R}_{0}}{R} \right)^5 + \left( \frac{{R}_{0}}{R} \right)^2 - \frac{2R}{{R}_{0}}  \right] .
\end{equation}
Next, collecting the $O(a)$ terms, we obtain the governing, second-order differential equation for the bubble-wall perturbation magnitude, $a$:
\begin{equation}
a_{,tt} + \eta a_{,t}  - \xi a = 0\,; \quad \eta = \frac{3 R_{,t}}{R} + \frac{4 \mu}{ \rho R^2} + \frac{ l(l+1) \mu}{ 3 \rho R^2}, \label{eq:a_sec_ode}
\end{equation}
where $\xi$  explicitly accounts for inertial effects during cavitation,  nonlinear deformations of the viscoelastic solid, and  surface tension effects: 
\begin{equation}
\small
    \begin{aligned}
    \xi = & - \frac{R_{,tt}}{R} + \frac{4 \mu R_{,t}}{\rho R^3} - \frac{ 2 l(l+1) \mu R_{,t}}{ 3 \rho R^3}  - \frac{(l+2)(l-1)\gamma}{\rho R^3}
     - \frac{2 G R_{0} }{\rho R^3} \Big( 1 + \frac{R_{0}^3}{R^3} \Big) - \frac{G l (l+1)}{\rho (R^2 + R R_{0} + R_{0}^2)} \\
    & - \frac{2 \alpha G}{\rho R^2} \frac{(R-R_{0})^2}{R R_{0}} \Big( 1+ \frac{R_{0}}{R} \Big)^3 \Big( 2-  \frac{2R_{0}}{R} +  \frac{3R_{0}^2}{R^2} - \frac{R_{0}^3}{R^3} + \frac{R_{0}^4}{R^4} \Big) 
    - \frac{\alpha G l(l+1) (R - R_{0})^2}{5 \rho R R_{0} (R^2 + R R_{0} + R_{0}^2)} \Big( 10 +  \frac{6R_{0}}{R} +  \frac{3R_{0}^2}{R^2} + \frac{R_{0}^3}{R^3} \Big) .
    \end{aligned} \label{eq:a_sec_ode_xi}
\end{equation}
We note that the perturbation of the pressure-like quantity $p$ does not enter the derivation of Eqs.\,\eqref{eq:RP}-\eqref{eq:a_sec_ode_xi}.

\section{Eigenvalues of Eq (7) in the main text}
The second order ordinary differential equation for $a(t)$, Eq.\,(7) in the main text, can be rewritten as the following dynamical system: 
\begin{equation}
    \frac{d}{dt} 
    \begin{bmatrix}
    a \\ 
    a_{,t}
    \end{bmatrix} = \begin{bmatrix}
    0 & 1 \\
    \xi & - \eta
    \end{bmatrix}
    \begin{bmatrix}
    a \\ 
    a_{,t}
    \end{bmatrix}. \label{eq:a_dynsys}
\end{equation}
It is straightforward to see that $\mathbf{a}^{*}$\,$:=$\,$\lbrace a=0, a_{,t}=0 \rbrace$ is always an equilibrium point of \eqref{eq:a_dynsys}, and its stability depends on the sign of the eigenvalues of the dynamical system \eqref{eq:a_dynsys}. The eigenvalues are of the form
    \begin{equation}
        \zeta_{1,2}= -\frac{\eta}{2} \pm \frac{\sqrt{\Delta}}{2}\quad\text{with}\quad\Delta = \eta^2 + 4 \xi, \label{eq:case_i}
    \end{equation}
which may be classified in the following cases:
\begin{itemize}
    \item \textbf{Case (i)}: If $\xi$\,$>$\,$0$, the eigenvalues are real and of opposite sign, and therefore, $\mathbf{a}^{*}$ is a \emph{saddle} point (Fig.\,\ref{fig:phase_diag} (a-i));
    \item \textbf{Case (ii)}: If $\xi$\,$<$\,$0$, $\eta<0$, and $\Delta > 0$, the eigenvalues are real and positive, and $\mathbf{a}^{*}$ is a \textit{source}-type instability point (Fig.\,\ref{fig:phase_diag} (a-ii));
    \item \textbf{Case (iii)}: If $\xi$\,$<$\,$0$, $\eta<0$, and $\Delta < 0$, the eigenvalues are complex with positive real part, and $\mathbf{a}^{*}$ is an \textit{unstable spiral}-type instability (Fig.\,\ref{fig:phase_diag} (a-iii));
    \item \textbf{Case (iv)}: If $\xi$\,$<$\,$0$, $\eta>0$, and $\Delta < 0$, the eigenvalues are complex with negative real part, and $\mathbf{a}^{*}$ is a \textit{stable spiral} (Fig.\,\ref{fig:phase_diag} (a-iv));
    \item \textbf{Case (v)}: If $\xi$\,$<$\,$0$, $\eta>0$, and $\Delta > 0$, the eigenvalues are real and negative, and  $\mathbf{a}^{*}$ is a \textit{sink} (Fig.\,\ref{fig:phase_diag} (a-v)).
\end{itemize}
A phase diagram summarizing the stability of the solution $\mathbf{a}^{*}$\,$=$\,$\lbrace a=0, a_{,t}=0 \rbrace$ for an inertial cavitation bubble in a nonlinear viscoelastic medium (Eq\,(7) in the main text) is shown in Fig.\,\ref{fig:phase_diag}(a).

%%%%%%%%%%%%%%%%%%%%%%%%%%%%%%%%%%%%%%%%%%%%%%%%%%%%%%
\section{Additional results of the numerical simulation in Fig 3}
\begin{figure}[h]
\centering
\includegraphics[width = 1 \textwidth]{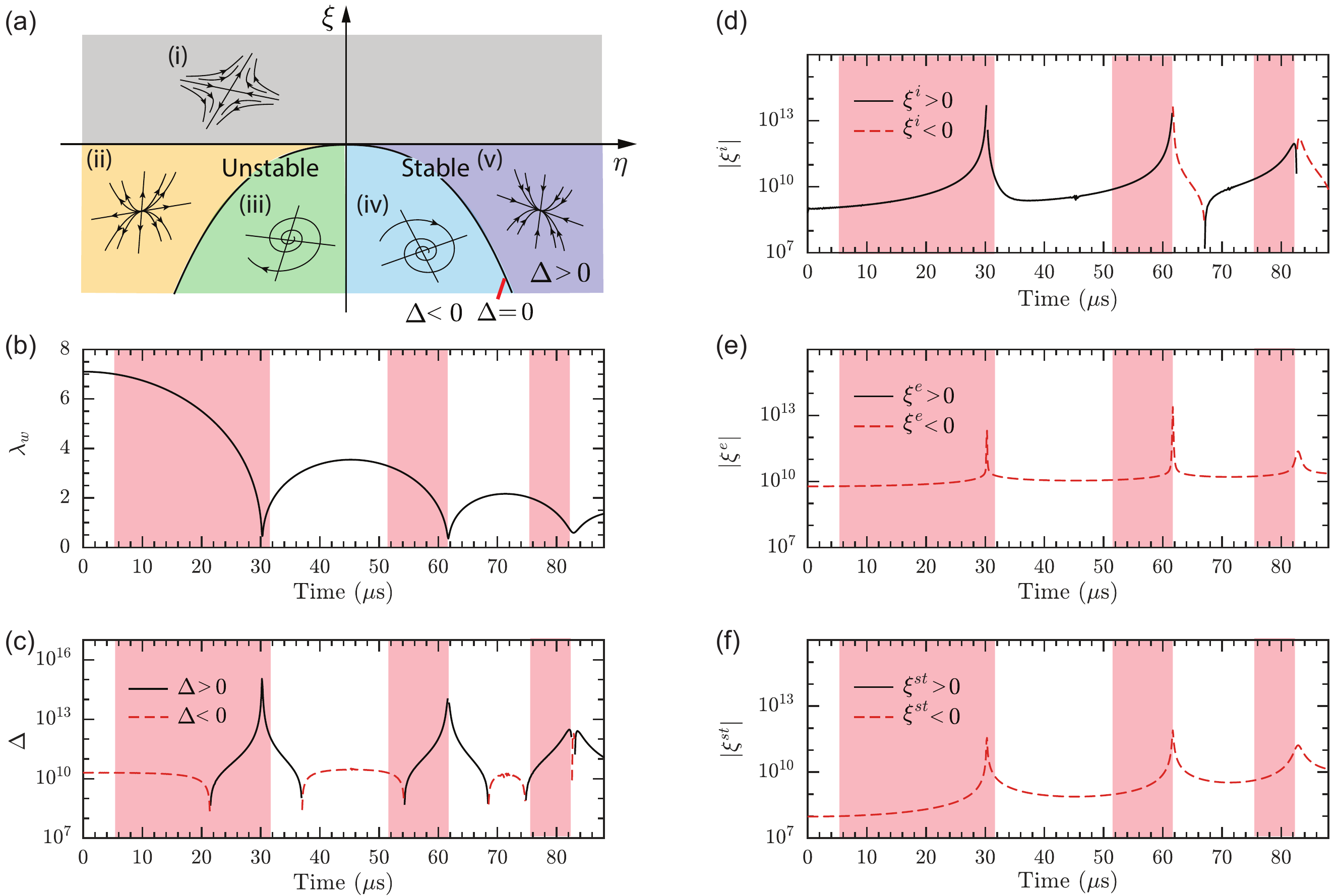}
\caption{(a)\,Phase diagram  summarizing the stability of the solution $\mathbf{a}^{*}$\,$=$\,$\lbrace a=0, a_{,t}=0 \rbrace$ for an inertial cavitation bubble in a nonlinear viscoelastic medium. Calculated (b)\,radial stretch ratio at the bubble wall $\lambda_w$, (c)\,$\Delta$, (d)\,$\xi^{i}$, (e)\,$\xi^{e}$, (f)\,$\xi^{st}$ in the numerical simulation in Fig.\,3 of the main text. Red shaded regions denote situations in which non-spherical instabilities corresponding to a spherical harmonic function of mode shape 8 are predicted.} \label{fig:phase_diag}
\end{figure}

Here we provide additional details of the numerical simulations in Fig.\,3 of the main text.  We summarize the numerically simulated hoop stretch ratio at the bubble wall $\lambda_w$ and the variables $\Delta$, $\xi^{i}$, $\xi^{e}$, and $\xi^{st}$ in Fig.\,\ref{fig:phase_diag}(b-f), where the variables $\xi^{i}$, $\xi^{e}$, $\xi^{st}$ account for  inertial and viscous effects during cavitation,  nonlinear hyperelastic deformations of the surrounding solid, and  surface tension effects, respectively:
\begin{align}
    & \xi^{i} = - \frac{R_{,tt}}{R} + \frac{4 \mu R_{,t}}{\rho R^3} + \frac{ 2 l(l+1) \mu R_{,t}}{ 3 \rho R^3},  \\
    & \begin{aligned} 
    \xi^{e} = & - \frac{2 G R_{0} }{\rho R^3} \Big( 1 + \frac{R_{0}^3}{R^3} \Big) - \frac{G l (l+1)}{\rho (R^2 + R R_{0} + R_{0}^2)} \\
    & - \frac{2 \alpha G}{\rho R^2} \frac{(R-R_{0})^2}{R R_{0}} \Big( 1+ \frac{R_{0}}{R} \Big)^3 \Big( 2-  \frac{2R_{0}}{R} +  \frac{3R_{0}^2}{R^2} - \frac{R_{0}^3}{R^3} + \frac{R_{0}^4}{R^4} \Big)  \\
    & - \frac{\alpha G l(l+1) (R - R_{0})^2}{5 \rho R R_{0} (R^2 + R R_{0} + R_{0}^2)} \Big( 10 +  \frac{6R_{0}}{R} +  \frac{3R_{0}^2}{R^2} + \frac{R_{0}^3}{R^3} \Big) ,
    \end{aligned} \\
    & \xi^{st} = - \frac{(l+2)(l-1)\gamma}{\rho R^3}.
\end{align}
Conditions in which non-spherical instabilities corresponding to a spherical harmonic function of mode shape 8 are predicted (i.e., $\eta<0$ or $\xi>0$) are marked as red shaded regions. % By further considering the phase space variable $\Delta$\,$=$\,$\eta^2$\,$+$\,$ 4 \xi$ (see Fig.\,\ref{fig:phase_diag}(c)), we find that the spatiotemporal evolution of this particular mode, i.e., mode 8, shows moderate growth in time, characterized by an \textit{unstable spiral} type instability first and then a \textit{saddle point} instability in the phase diagram.

%\vspace{0.75cm}
%%%%%%%%%%%%%%%%%%%%%%%%%%%%%%%%%%%%%%%%%%%%%%%%%%%%%%
%%%%%%%%%%%%%%%%%%%%%%%%%%%%%%%%%%%%%%%%%%%%%%%%%%%%%% 

%%%%%%%%%%%%%%%%%%%%%%%%%%%%%%%%%%%%%%%%%%%%%%
\section{Details of theoretical predictions in Fig 4}\label{sec:eee}
\begin{figure} [h!]
\begin{center} 
\includegraphics[width= 1 \textwidth]{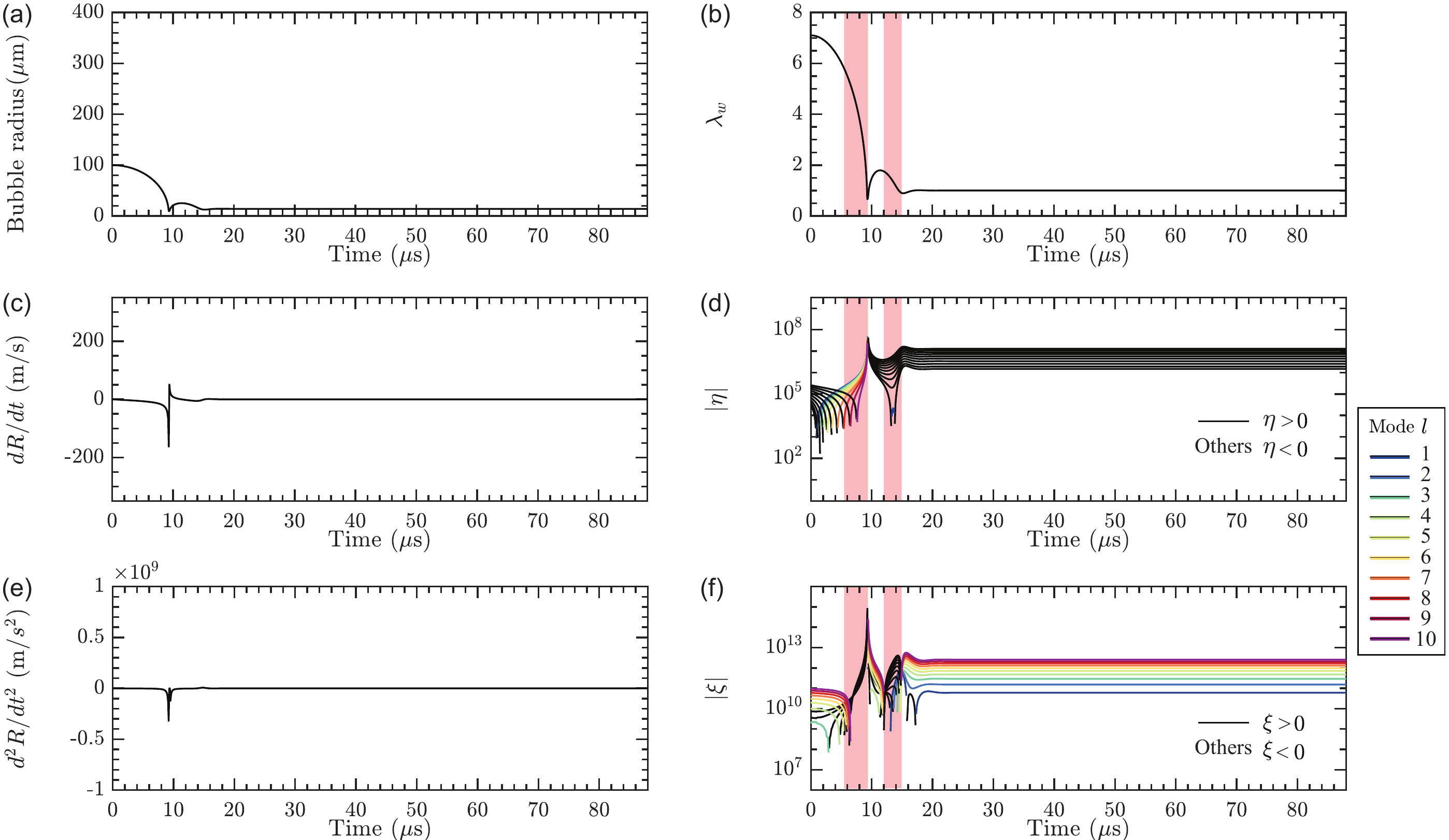}
\end{center}
\caption{(a,c,e)\,Numerically simulated bubble dynamics when $R_{\text{max}}=100\mu m$. (b,d,f)\,Calculated bubble wall stretch ratio $\lambda_w$ and the variables $\eta$ and $\xi$ calculated using Eqs\,(7-8) in the main text. Red shaded regions denote theoretically predicted  non-spherical mode shape 8 instabilities. } \label{fig:R100}
\end{figure}

\begin{figure} [h!]
\begin{center} 
\includegraphics[width= 1 \textwidth]{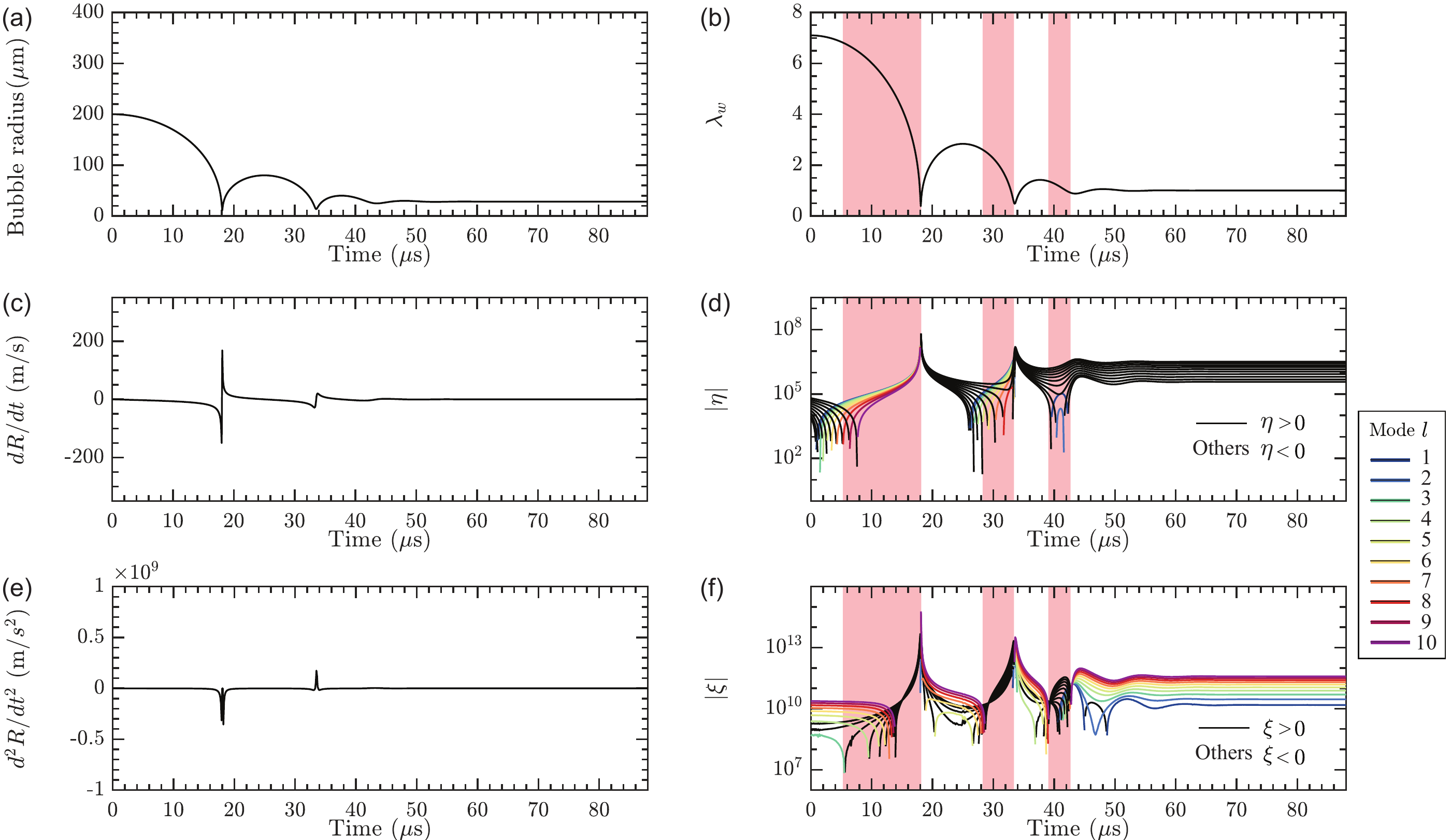}
\end{center}
\caption{(a,c,e)\,Numerically simulated bubble dynamics when $R_{\text{max}}=200\mu m$. (b,d,f)\,Calculated bubble wall stretch ratio $\lambda_w$ and the variables $\eta$ and $\xi$ calculated using Eqs\,(7-8) in the main text.  Red shaded regions denote theoretically predicted  non-spherical mode shape 8 instabilities.} \label{fig:R200}
\end{figure}

\begin{figure} [h!]
\begin{center} 
\includegraphics[width= 1 \textwidth]{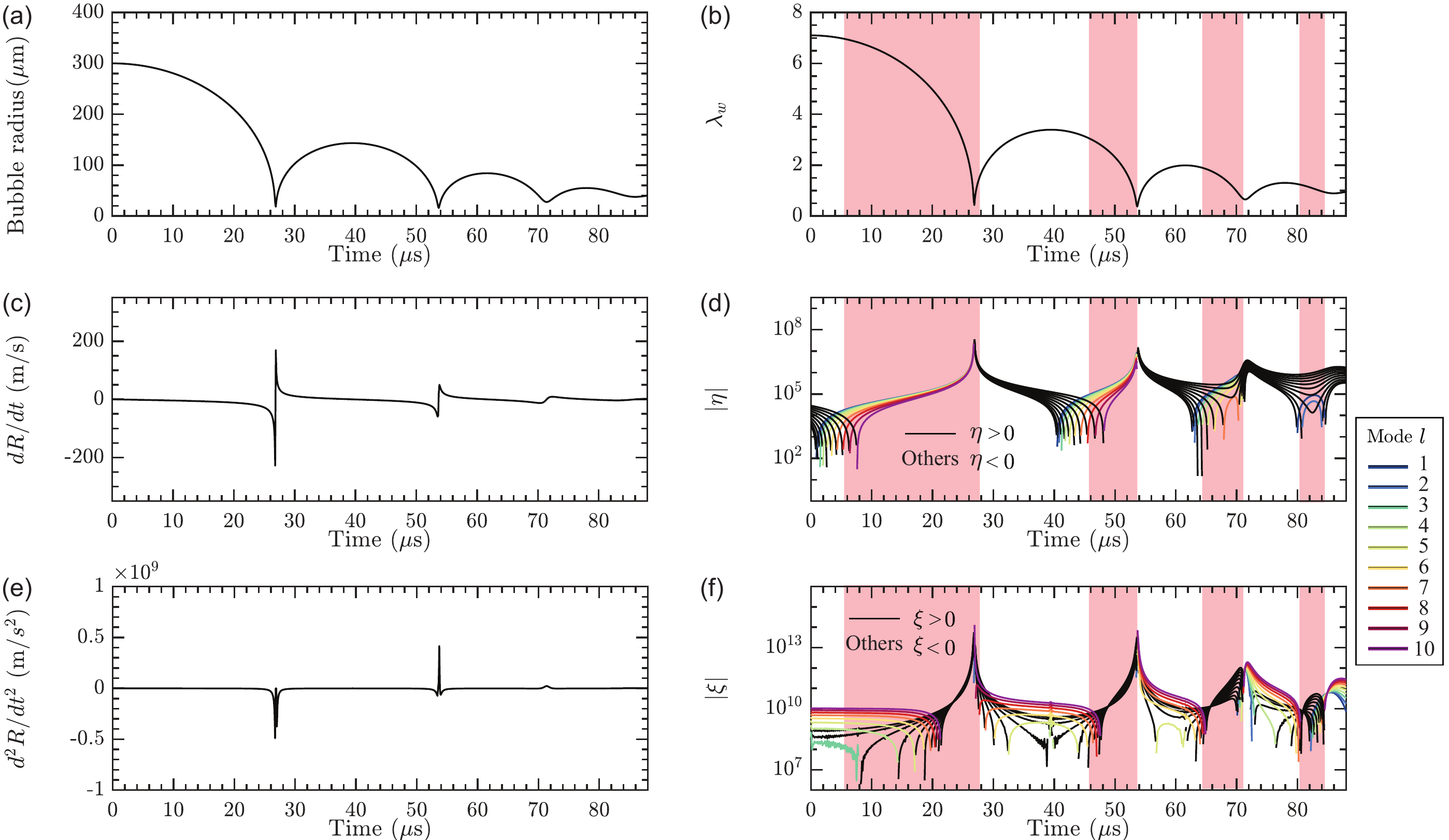}
\end{center}
\caption{(a,c,e)\,Numerically simulated bubble dynamics when $R_{\text{max}}=300\mu m$. (b,d,f)\,Calculated bubble wall stretch ratio $\lambda_w$ and the variables $\eta$ and $\xi$ calculated using Eqs\,(7-8) in the main text.  Red shaded regions denote theoretically predicted  non-spherical mode shape 8 instabilities.} \label{fig:R300}
\end{figure}

\begin{figure} [h!]
\begin{center} 
\includegraphics[width= 1 \textwidth]{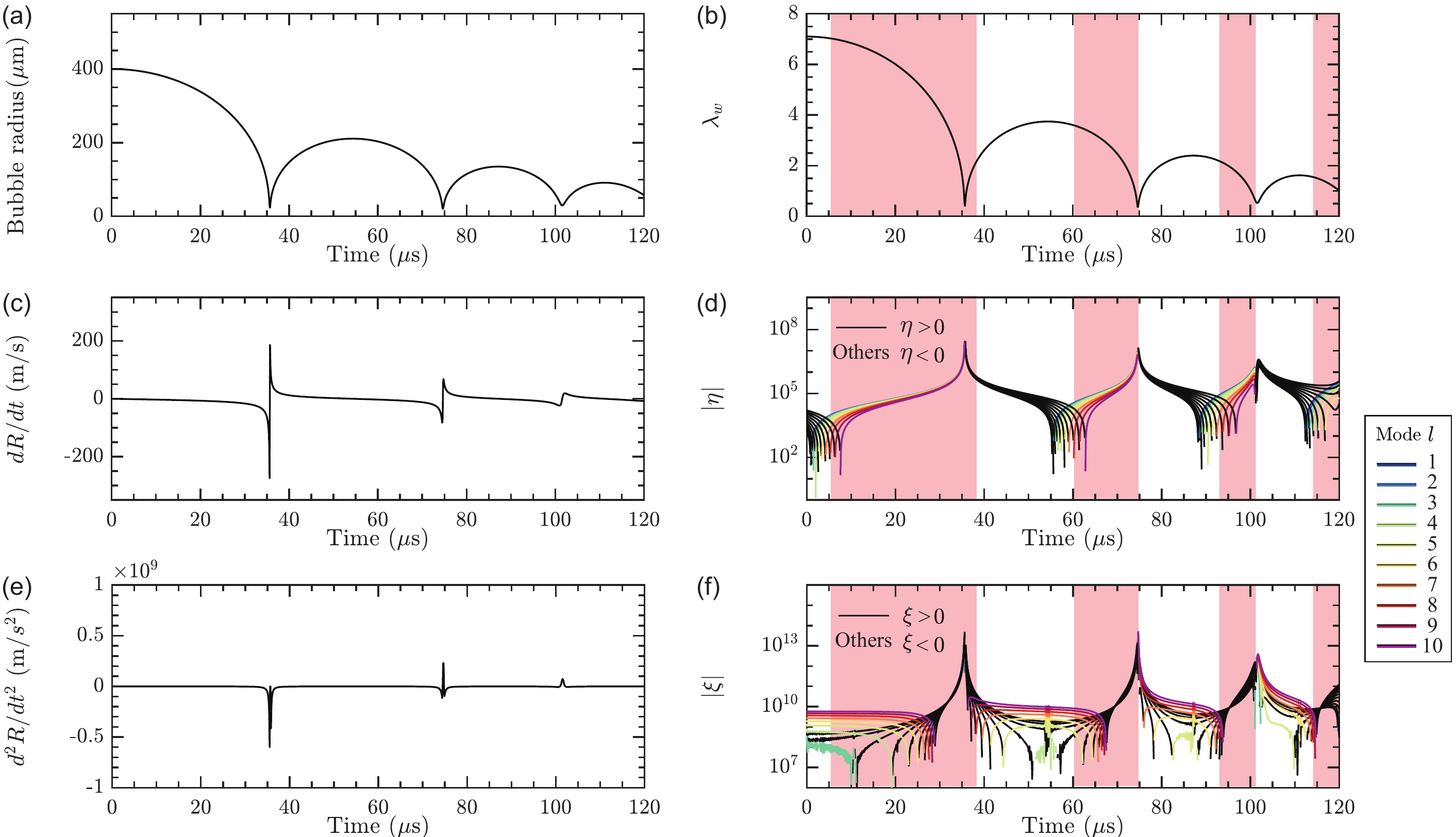}
\end{center}
\caption{(a,c,e)\,Numerically simulated bubble dynamics when $R_{\text{max}}=400\mu m$. (b,d,f)\,Calculated bubble wall stretch ratio $\lambda_w$ and the variables $\eta$ and $\xi$ calculated using Eqs\,(7-8) in the main text.  Red shaded regions denote theoretically predicted  non-spherical mode shape 8 instabilities.} \label{fig:R400}
\end{figure}

\begin{figure} [h!]
\begin{center} 
\includegraphics[width= 1 \textwidth]{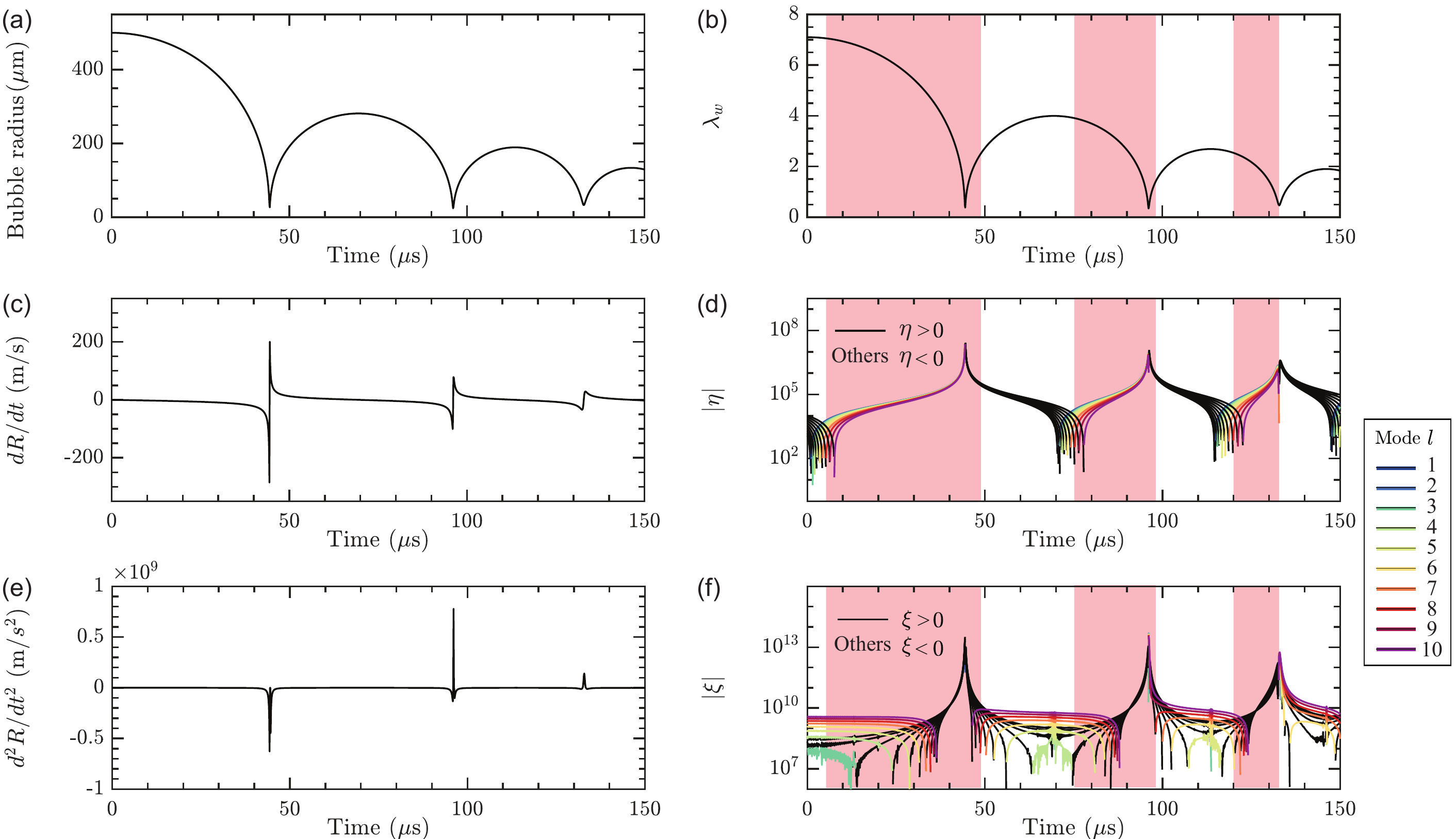}
\end{center}
\caption{(a,c,e)\,Numerically simulated bubble dynamics when $R_{\text{max}}=500\mu m$. (b,d,f)\,Calculated bubble wall stretch ratio $\lambda_w$ and the variables $\eta$ and $\xi$ calculated using Eqs\,(7-8) in the main text.  Red shaded regions denote theoretically predicted  non-spherical mode shape 8 instabilities.} \label{fig:R500}
\end{figure}

%%%%%%%%%%%%%%%%%%%%%%%%%%%%%%%%%%%%%%%%%%%%%%
We theoretically predict the critical value of the radial stretch ratio at the bubble wall, $\lambda_w$, at the onset of bubble surface instability during the first three collapse cycles as a function of the  maximum bubble radius $R_{\text{max}}$ and the non-spherical mode shape number $l$. The initial $\lambda_{\text{max}}$ is fixed at 7.1, which is the same value as in all experimental tests  shown in Figs.\,1-2 in the main text.  
For each $R_{\text{max}}$ value, we numerically solve the Rayleigh-Plesset equation, Eqs.\,(1) and (3), in the main text and compute the variables $\eta$ and  $\xi$ using Eqs\,(7-8) in the main text. Then, we find the critical $\lambda_{w}$ at the onset of each  instability mode occurring in the first three collapse cycles where $\eta <0$ or $\xi>0$, as shown in Figs.\,\ref{fig:R100}-\ref{fig:R500}, where the theoretically predicted  non-spherical mode shape 8 instabilities are marked as red shaded regions. In Figs.\,\ref{fig:R100}-\ref{fig:R500}, $R_{\text{max}}$ takes a value of 100$\mu m$, 200$\mu m$, 300$\mu m$, 400$\mu m$, 500$\mu m$, respectively.

%%%%%%%%%%%%%%%%%%%%%%%%%%%%%%%%%%%%%%%%%%%%%%
%%%%%%%%%%%%%%%%%%%%%%%%%%%%%%%%%%%%%%%%%%%%%%

%%%%%%%%%%%%%%%%%%%%%%%%%%%%%%%%%%%%%%%%%%%%%%
%%%%%%%%%%%%%%%%%%%%%%%%%%%%%%%%%%%%%%%%%%%%%%

%%%%%%%%%%%%%%%%%%%%%%%%%%%%%%%%%%%%%%%%%%%%%%
%%%%%%%%%%%%%%%%%%%%%%%%%%%%%%%%%%%%%%%%%%%%%%

%%%%%%%%%%%%%%%%%%%%%%%%%%%%%%%%%%%%%%%%%%%%%%
%%%%%%%%%%%%%%%%%%%%%%%%%%%%%%%%%%%%%%%%%%%%%%

%%%%%%%%%%%%%%%%%%%%%%%%%%%%%%%%%%%%%%%%%%%%%%
%%%%%%%%%%%%%%%%%%%%%%%%%%%%%%%%%%%%%%%%%%%%%%

%%%%%%%%%%%%%%%%%%%%%%%%%%%%%%%%%%%%%%%%%%%%%%
\section{Finite element Numerical Simulations}\label{sec:fem}
%%%%%%%%%%%%%%%%%%%%%%%%%%%%%%%%%%%%%%%%%%%%%%
To  examine the strain and stress fields induced by non-spherical deformation, we construct a finite-element model of an isolated 3D spherical cavity within an infinite material using the commercial software package Abaqus \cite{abaqus}. Using a user-material (UMAT) subroutine, we model the surrounding material as a quadratic law strain stiffening Kelvin-Voigt (qKV) viscoelastic material (Eq\,(2) in the main text) with shear modulus $G$\,$=$\,$2.77$ kPa, strain stiffening  parameter $\alpha$=0.48, and viscosity $\mu$\,$=$\,$0.186$\,Pa$\cdot$s. %To model the infinite solid, we developed a UMAT subroutine to account for large strain viscoelasticity, where we applied the quadratic strain stiffening Kelvin-Voigt model (Eq\,(2) in the main text) with shear modulus $G$\,$=$\,$2.77$ kPa, strain stiffening  parameter $\alpha$ 0.48, and viscosity $\mu$\,$=$\,$0.186$\,Pa$\cdot$s. 
Due to the symmetry of the problem, we simulate one eighth of the sphere. The undeformed bubble radius is taken to be 45\,$\mu$m, and the outer radius of the simulation domain is taken to be large enough so that an infinite surroundings is approximated. The mesh consists of 7275 3D, 8-node fully integrated hybrid elements. Next, to account for the non-spherical deformation, we impose a displacement boundary condition on the bubble wall surface consisting of the sum of a spherically symmetric deformation and a spherical harmonic perturbation (Eq.\,(4) in the main text), using a user-displacement (UDISP) subroutine. We perform a static simulation considering a time point between (iv) and (v) in the main text Fig.\,1(a), corresponding to a deformed bubble radius of 72\,$\mu$m, a perturbation amplitude of 1.6386 $\mu$m, and mode shape $l=8$. %the experimentally fitted perturbed displacement boundary condition (Eq\,(4) in the main text) at the boundary between the cavity and the infinite material using a user-displacement (UDISP) subroutine. We performed a static simulation considering time points from (iv) to (v) as seen in the main text Fig.\,1(a). We assigned the equilibrium radius to be equal to the experimentally measured radius which is around 45 $\mu$m and for the infinite material we set a radius large enough to resolve any pressure waves emitted during the collapse phase of the spherical cavity. The mesh finally consisted of 7275 of 3D, 8-noded fully integrated hybrid elements.
 
\begin{figure} [h!]
\begin{center} 
\includegraphics[width= 0.6  \textwidth]{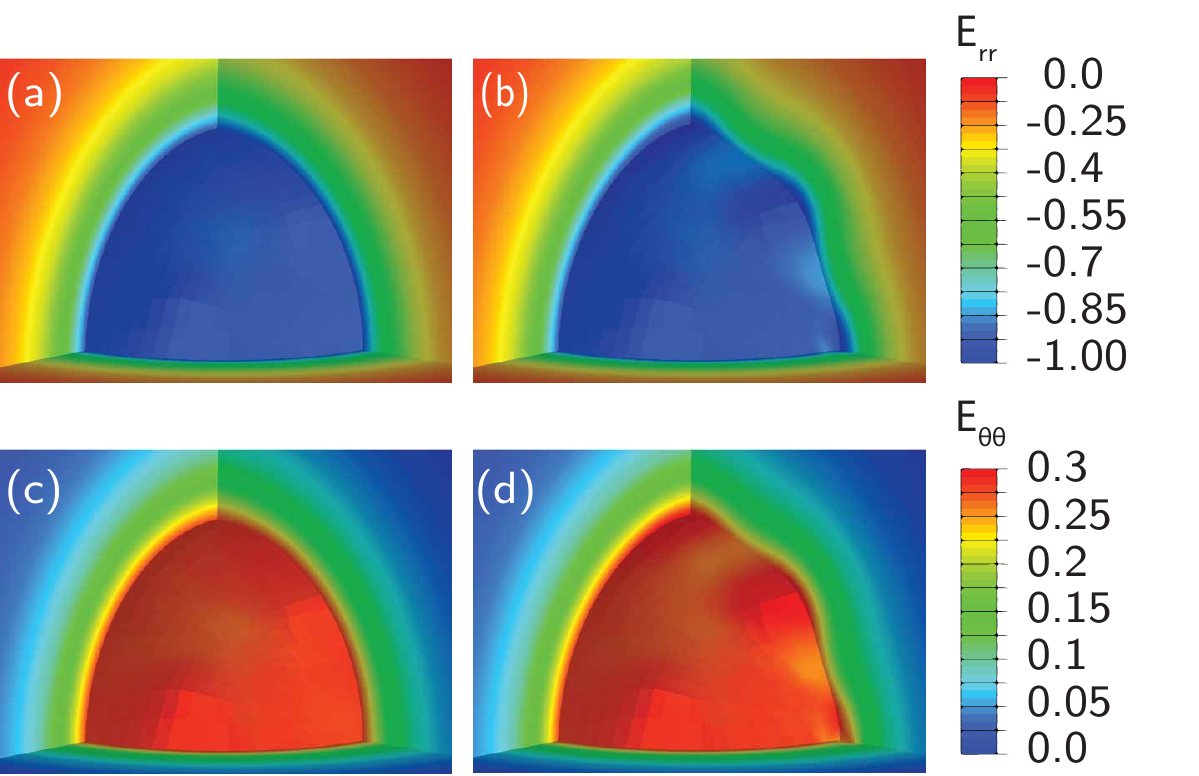}
\end{center}
\caption{Contours of the spherical component of the logarithmic strain, $ E_{rr}$, for (a) the spherical bubble and (b) the perturbed bubble. Contours of the circumferential component of the logarithmic strain, $ E_{\theta \theta}$, for (c) the spherical bubble and (d) the perturbed bubble. The deformed radius of the spherical bubble is 72 $\mu$m, and the perturbation amplitude in (b) and (d) is 1.6386 $\mu$m.} \label{fig:strain contour plots}
\end{figure}

\begin{figure} [h!]
\begin{center} 
\includegraphics[width= 0.6  \textwidth]{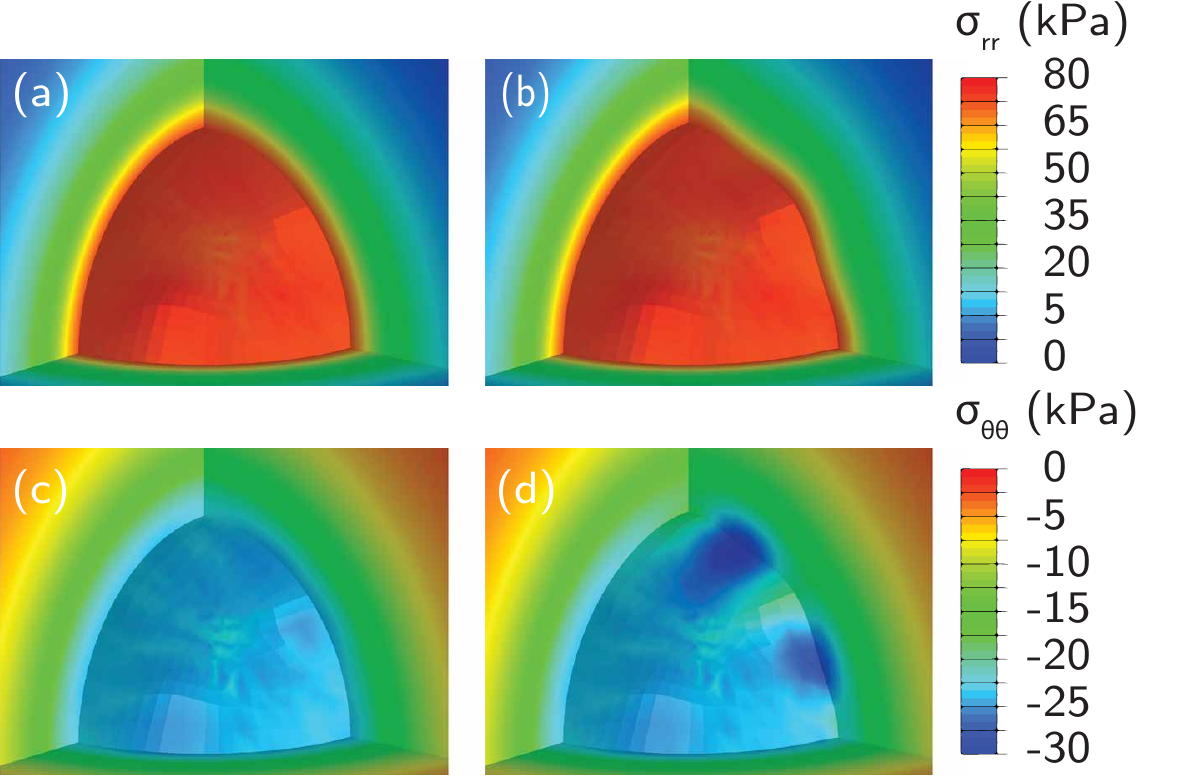}
\end{center}
\caption{Contours of the spherical component of the Cauchy stress, $\sigma_{rr}$, for (a) the spherical bubble and (b) the perturbed bubble. Contours of the circumferential component of the Cauchy stress, $\sigma_{\theta \theta}$, for (c) the spherical bubble and (d) the perturbed bubble. The deformed radius of the spherical bubble is 72 $\mu$m, and the perturbation amplitude in (b) and (d) is 1.6386 $\mu$m.} \label{fig:stress contour plots}
\end{figure}

In Fig.\,\ref{fig:strain contour plots}(a) and (b), we plot contours of the spherical component of the logarithmic strain, $ E_{rr}$, while in Fig.\,\ref{fig:strain contour plots}(c) and (d) we plot contours of the circumferential component of the logarithmic strain $ E_{\theta \theta}$.  
In Fig.\,\ref{fig:stress contour plots}(a) and (b) we plot contours of the spherical component of the Cauchy stress, $\sigma_{rr}$, while in Fig.\,\ref{fig:stress contour plots}(c) and (d) we plot contours of the circumferential component of the Cauchy stress, $\sigma_{\theta \theta}$.  We include both the spherically symmetric case and the perturbed case, illustrating the amplification of stress and strain due to non-spherical deformation, notably in the circumferential components of stress and strain. 
 
%%%%%%%%%%%%%%%%%%%%%%%%%%%%%%%%%%%%%%%%%%%%%%
%%%%%%%%%%%%%%%%%%%%%%%%%%%%%%%%%%%%%%%%%%%%%%

%%%%%%%%%%%%%%%%%%%%%%%%%%%%%%%%%%%%%%%%%%%%%%
%%%%%%%%%%%%%%%%%%%%%%%%%%%%%%%%%%%%%%%%%%%%%%

\bibliography{reference}